%% file: ms.tex
\documentclass[12pt,preprint]{aastex}

\slugcomment{Feb. 16}

\shorttitle{Near-Infrared Colors of Hard X-ray Selected AGNs}
\shortauthors{Watanabe et al.}

\begin{document}

\title{Near-Infrared Colors of Hard X-ray Selected Active Galactic
Nuclei\altaffilmark{1}
  } 

\author{Chisato Watanabe
\altaffilmark{2}
}
\affil{Department of Astronomy,
Kyoto University, Kyoto, 606-8502, Japan}
\email{chisato@kusastro.kyoto-u.ac.jp}

\author{Kouji Ohta
\altaffilmark{3}
}

\affil{Department of Astronomy,
Kyoto University, Kyoto, 606-8502, Japan}
\email{ohta@kusastro.kyoto-u.ac.jp}

\author{Masayuki Akiyama
\altaffilmark{2,3}
}

\affil{Subaru Telescope, National Astronomical Observatory of Japan,
Hilo, HI, 96720}
\email{akiyama@subaru.naoj.org}

\and

\author{Yoshihiro Ueda}
\affil{Institute of Space and Astronautical Science, Sagamihara,
Kanagawa, 229-8510, Japan}
\email{ueda@astro.isas.ac.jp}

\altaffiltext{1}{Based on observations made with the Kitt Peak National
Observatory 2.1m telescope, which is operated by National Optical
Astronomy Observatories (NOAO) operated by AURA, Inc., under contract
with the National Science Foundation and with the University of Hawaii 
88 Inch telescope. }
\altaffiltext{2}{Visiting Astronomer, University of Hawaii Observatory.}
\altaffiltext{3}{Visiting Astronomer, Kitt Peak National
Observatory, National Optical Astronomy Observatories (NOAO).}

\begin{abstract}
We present results of near-infrared photometry ($J$, $H$, $K_S$) 
for a hard X-ray selected sample of active galactic nuclei (AGNs) 
obtained from optical 
identification of the sources detected in {\it ASCA} 
surveys (total $\sim$75 deg$^2$) with a flux limit of 
$(1-3) \times 10^{-13}$ erg s$^{-1}$ cm$^{-2}$ (2--10 keV). 
The sample covers the AGNs at 0.1 $\lesssim z \lesssim$ 1 
with $L_{\rm 2-10keV} \sim 10^{42} - 10^{46}$ erg s$^{-1}$ 
with very high completeness. 
The near-infrared photometric data of the sample are obtained from 
the Two Micron All Sky Survey (2MASS) and observations with 
Kitt Peak National Observatory 2.1 m telescope and 
the University of Hawaii 88 inch telescope. 
The fraction of red ($J-K_{S}>2$ mag) AGNs in our sample 
is $\sim 2\pm1\%$, which is comparable to that for 
optically- or UV-selected quasi-stellar objects 
(QSOs, i.e. luminous AGNs). 
The number of red AGNs found in our sample is consistent 
with that expected from the surface density of red 
AGNs found in 2MASS by Cutri et al. (2001). 
We find that the anomalously-small dust-to-gas ratios 
in circumnuclear gas, which is seen 
in some AGNs with Seyfert-class luminosity, 
also occur in the QSOs 
(AGNs with luminosity of $L_{\rm 2-10keV} \ge 10^{44.5}$ erg s$^{-1}$) 
in the systematically surveyed sample in hard X-rays. 
For all the QSOs with an X-ray absorption of 
$N_{\rm H} \ge 10^{22}$ cm$^{-2}$ in our sample, 
the values of $A_V/N_{\rm H}$
are smaller than the Galactic value by a factor of $\sim$5 to $\sim$100. 
Since a fraction of this population among the QSOs in our sample 
is about 30\%, such fraction of optical/UV-selected
type 1 QSOs known to date may show type 2 nature in X-ray. 

\end{abstract}


\section{Introduction}

Many candidates of red, presumably moderately absorbed, quasi-stellar
objects (QSOs) --- luminous active galactic nuclei (AGNs) --- 
have recently been found in QSO
surveys in radio, X-ray, and near-infrared (NIR) wavelengths.
Webster et al. (1995) revealed the presence of many
red QSOs among radio-selected QSOs and suggested that
the red color originated in an absorption to its nucleus.
Since a significant fraction of the radio-selected QSOs has 
redder colors than optically-selected QSOs
($B-K=2\sim3$ mag), it is argued that 80\% of QSO population could be
significantly absorbed and escape from traditional optical/UV
selections.
Subsequent studies revealed that
the origin of the red color is not absorption to
its nucleus, but a red synchrotron continuum of a
jet component in $\sim 80$\% of the red radio-loud QSOs 
(Francis et al. 2001).
However, it is still possible that there is a large 
population of moderately ($E(B-V) \lesssim 1$ mag) reddened QSOs 
in radio-selected QSOs (White et al. 2003). 

On the other hand, for radio-quiet QSOs, which are the dominant
population of QSOs in the optical and X-ray wavelengths, the 
existence of red broad-line QSOs is reported  in identification of
soft X-ray selected {\it ROSAT} sources (Kim \& Elvis 1999).
Since the QSOs have harder X-ray spectra than optically-selected
QSOs and  have large H$\alpha$ to H$\beta$ broad-line flux ratios,
it is likely that the extinction of $A_V \sim 2$ mag towards the nucleus  
makes the colors of the QSOs redder than optically-selected QSOs.
The fraction of the red ($B-R>2$ mag) QSO is not large ($\sim 1\%-7\%$)
in the soft X-ray selected sample, but a number of absorbed
QSOs could be equal to that of the blue optically-selected QSOs
(Kim \& Elvis 1999), because the soft X-ray selection is biased 
to less absorbed QSOs and heavily absorbed QSOs are hard to be detected 
in the soft X-ray band. 
More recently, in the NIR wavelength, Two Micron All Sky Survey (2MASS) 
also has revealed the presence of many AGNs  
with NIR colors redder than $J-K_S=2$ mag
at $z\lesssim0.3$ (2MASS red AGNs). 
The number density of the red AGNs is claimed to be comparable to 
that of optically-selected AGNs with $K_S < 14.5$ mag (Cutri et al.
2001, 2002).

The discoveries of red AGN population suggest that we have been missing
a significant fraction of AGNs in traditional
optical/UV selections of AGNs due to the moderate 
absorption to the nucleus. 
However the fraction of the red absorbed AGNs in the entire AGN
population is not clear.
Radio-selected samples are affected by red QSOs with the red synchrotron
component.
The Soft X-ray selection is biased to less absorbed QSOs and thus
not suitable to disclose the fraction of the absorbed QSOs.
The 2MASS-selected red AGNs are limited in the low redshift 
universe, primary because of the effect of the $K$-correction on intrinsic 
NIR colors of QSOs. 
Using a hard X-ray selected sample is one of the best ways to examine it 
because the hard X-ray is less affected by absorption, 
although very recently Wilkes et al. (2002) and 
Risaliti et al. (2003) have shown the presence of X-ray faint QSOs. 

Another important issue in terms of absorption to an active nucleus is
anomaly of dust-to-gas ratio in a circumnuclear region.
In some of AGNs, an extinction to a nucleus  estimated in optical light
is significantly smaller than that derived from X-ray spectrum
(e.g., Simpson 1998; Akiyama et al. 2000a; Risaliti et al. 2001; 
Akiyama, Ueda, \& Ohta 2002b; Willott et al. 2003).
Since the former extinction is caused by dust, while the latter
by gas phase metal, this suggests that a  dust-to-gas ratio is
much smaller than the standard value obtained in the Galaxy.
Maiolino et al. (2001a) compiled Seyfert galaxies showing 
evidence for absorption in the X-ray and/or NIR/optical data
from literature,
and found that the $A_V/N_{\rm H}$ values of most of them are
10 to 100 times smaller than the Galactic value.
They claim that the most likely interpretation of this is the depletion of
small grains due to coagulation 
(Maiolino, Marconi, \& Oliva 2001b), though there may be
other possibilities such as dust depletion due to sublimation
by AGN radiation, or a difference in locations of X-ray and 
optical absorbing matters (Weingartner \& Murray 2002). 
In any cases,  the anomaly in dust-to-gas ratio
affects not only on the understanding of
physical conditions in a circumnuclear region, but also on 
the view of the unified scheme of AGNs; there may be a population
that shows type 1 nature in optical/UV region but show type 2
nature in X-ray region.
The sample by Maiolino et al. (2001a) is rather heterogeneous and
limited to low luminosity AGNs ($<10^{44.5}$ erg s$^{-1}$ in a
2--10 keV band).
We intend to extend the luminosity range higher (QSO regime)
using an unbiased systematic sample obtained through  optical follow-up of
hard-X-ray source surveys.

To make a sample of AGNs at a higher redshift 
with a larger luminosity without biasing to less absorbed objects, 
we constructed a highly complete hard X-ray selected AGN sample 
covering a large survey area. 
Using a 2--10~keV hard X-ray emission, we can detect AGNs with 
an absorption up to $N_{\rm H}$ of $10^{22.5}\sim10^{23}$ cm$^{-2}$ with 
a negligible bias in the redshift range below $\sim 1$. 
In fact, a {\it Beppo-SAX} hard X-ray survey found a 
red QSO (Vignali et al. 2000), and the presence of many red 
optical counterparts found in deep {\it Chandra} and/or
{\it XMM-Newton} surveys (Hasinger et al. 2001) hints that
there are many red AGNs.
However, the optical identification of these samples is not complete 
so it has been difficult to draw conclusions about what fraction 
of all AGN that is absorbed. 

In this paper, we present  NIR colors of 
hard X-ray (2--10 keV) selected AGN samples obtained
through optical identification of hard X-ray surveys with {\it ASCA}. 
The optical identification is fairly complete and 
the sample is suitable to estimate the fraction of 
red AGNs and the distribution of
$A_V/N_{\rm H}$ values in QSOs at intermediate redshifts.
Throughout this paper, we use cosmological parameters of 
$H_0=70$ km s$^{-1}$ Mpc$^{-1}$, $\Omega_m =0.3$, and 
$\Omega_\Lambda=0.7$. 

\section{Hard X-ray Selected Samples}
We use two hard X-ray selected samples of AGNs.
One is a sample from the {\it ASCA} Large Sky Survey (ALSS) and its optical
identification.
The ALSS is a contiguous ($\sim 5$ deg$^2$) sky survey near
the north Galactic pole made with {\it ASCA} during a period from 1993
and 1995 (Ueda et al. 1999a).
34 X-ray sources with a flux larger than $\sim 1 \times 10^{-13}$
erg s$^{-1}$ cm$^{-2}$ (2--10 keV) were found
in this survey (significance level larger than 3.5 $\sigma$
in the 2--7 keV Solid-state Imaging Spectrometer (SIS) band).
Optical follow-up spectroscopy was made and 33 sources were
successfully identified (Akiyama et al. 2000a);
they consist of 30 AGNs at redshifts of 0.03 to 1.7, two clusters of
galaxies and a Galactic star (one remaining X-ray source
is possibly a false source; see footnote of Ueda et al. (2003)).
The second sample is from the {\it ASCA} Medium
Sensitivity Survey (AMSS) and its optical identification.
The AMSS is a serendipitous source survey with {\it ASCA}; 
{\it ASCA} pointing observation data taken during a period from
1993 and 1996 were gathered and more than 1300 sources 
(including target sources) were detected 
in the fields of views of the Gas Imaging Spectrometer
(GIS) (Ueda et al. 1999b, 2001).
Akiyama et al. (2003) defined a complete flux-limited 
sample for optical follow-up by selecting sources in the 
northern sky with fluxes larger than 
$3 \times 10^{-13}$ erg s$^{-1}$ cm$^{-2}$ detected with 
significance above 5.5 $\sigma$ in the 2--10 keV band 
(AMSSn sample; more detailed criteria 
are described  by Akiyama et al. 2003). 
The AMSSn contains 87 objects in a total survey area of $\sim 70$ deg$^2$. 
We have identified all but one of
them; 78 sources are AGNs (including 3 BL Lac objects),
7 clusters and 1 Galactic star (Akiyama et al. 2003).

In total, 105 AGNs (excluding 3 BL Lac objects) are found in the ALSS
and AMSSn. The combined {\it ASCA} sample is unique even in comparison with
AGN samples from {\it Chandra} and {\it XMM-Newton} surveys, because it is the
largest hard X-ray selected AGN sample at a relatively bright flux
limit of 10$^{-13}$ erg s$^{-1}$ cm$^{-2}$ (2--10 keV). Table~1
summarizes the observed 2--10 keV flux and the absorption-corrected
luminosity in the rest-frame 2--10 keV band for
each {\it ASCA} source. We also list the intrinsic photon index $\Gamma$ and
absorption column-density to the nucleus $N_{\rm H}$ (and their
$1\sigma$ errors), which are used to calculate the flux and
luminosity. The photon index or absorption is estimated as 
described by Ueda et al. (2003), who utilize the hardness ratio
above and below 2 keV (or the 3-bin GIS spectrum plus 2-bin SIS
spectrum for the ALSS; see Ueda et al. 1999a for details); $N_{\rm
H}=0$ is assumed if the best-fit $\Gamma$ is larger than 1.9 while
$\Gamma=1.9$ for the other case. For AGNs for which follow-up
observations were made with {\it XMM-Newton} or {\it ASCA}, both $\Gamma$ and
$N_{\rm H}$ are determined through X-ray spectral fits. Note that
these values listed in Table~1 are slightly different from those in
Akiyama et al. (2000a, 2003), where $\Gamma=1.7$ is adopted (also, the
flux values in their plots are based on an unabsorbed power-law
spectrum with a varying photon index). The difference does not affect
on our overall discussion, however.

\begin{figure}
 \begin{center}
  \begin{tabular}{cc}
  \resizebox{80mm}{!}{\includegraphics{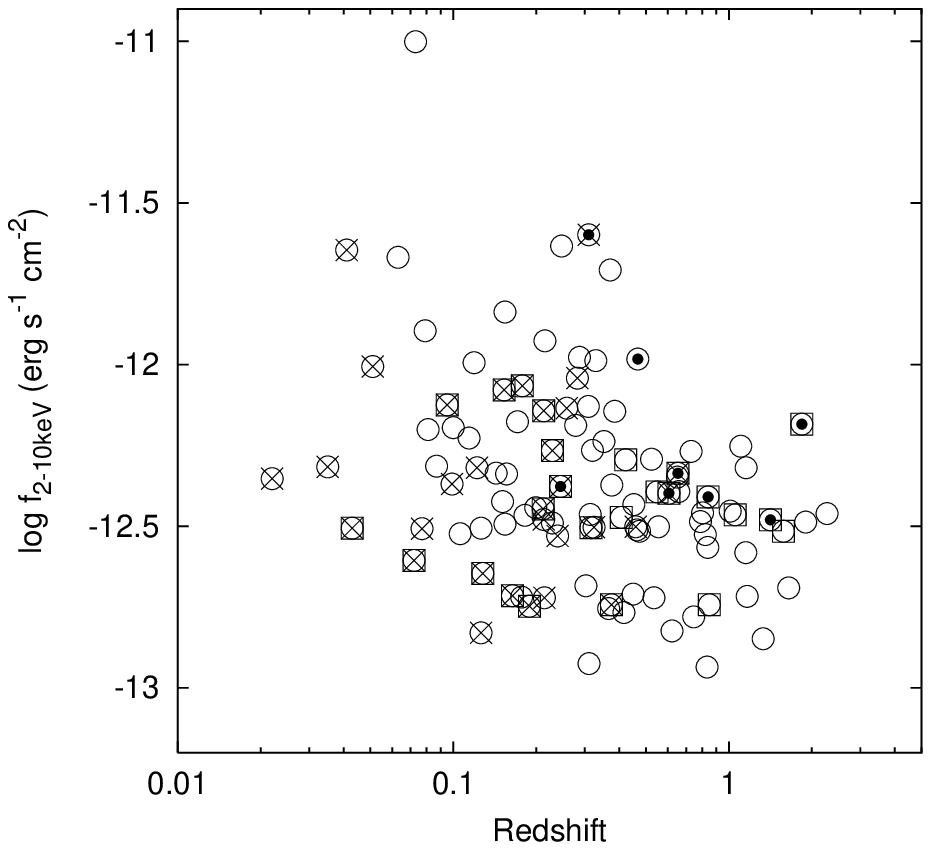}} & 
  \resizebox{80mm}{!}{\includegraphics{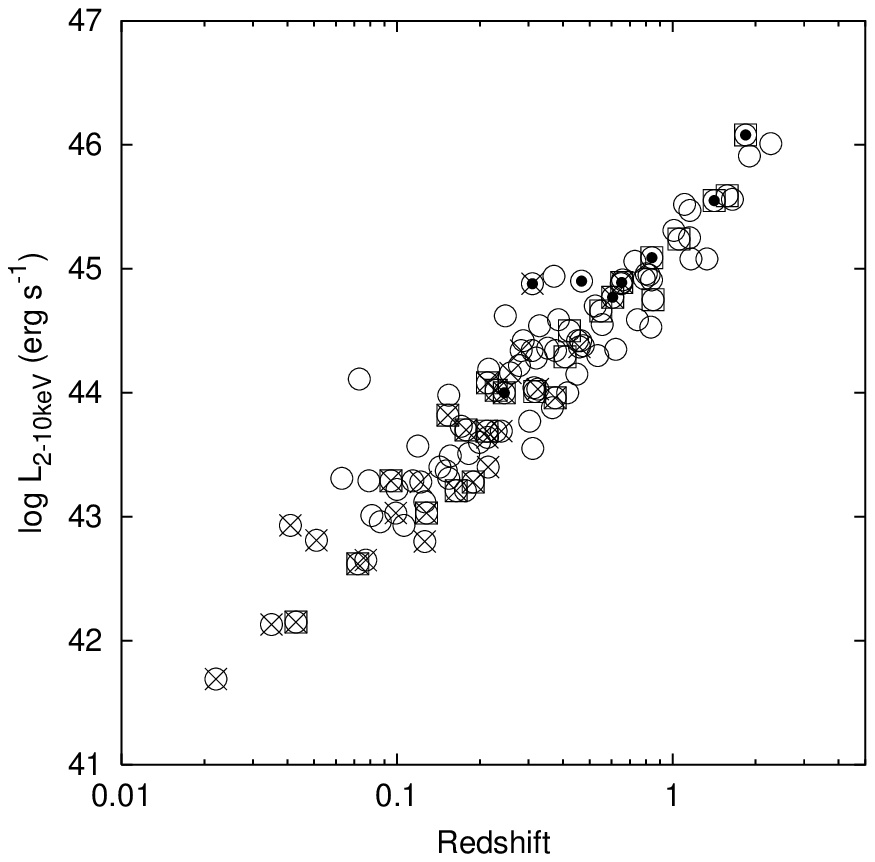}} \\
  \end{tabular}
 \end{center}
\caption{
$Left$: 2--10 keV flux against redshift for the {\it ASCA} AGNs. 
Over-plotted filled circles, crosses, and open squares 
indicate AGNs with a large hard X-ray to 
optical flux ratio (optically faint; log $f_{\rm 2-10~keV}/f_R > 1$), 
AGNs showing no significant broad H$\beta$ emission line, 
and AGNs with large X-ray absorption 
($N_{\rm H} \ge 10^{22}$ cm$^{-2}$), respectively.}
$Right$: Absorption corrected X-ray luminosity 
in the rest-frame 2--10 keV band
against redshift for the {\it ASCA} AGNs. 
\label{L_z}
\end{figure}

Figure \ref{L_z} shows the 2--10 keV flux against redshift for the 
sample in the left panel, and 
the absorption corrected luminosity 
in the rest-frame 2--10 keV band against redshift in the right panel. 
Over-plotted solid circles, crosses, and open squares 
indicate the AGNs with a large X-ray to optical 
flux ratio ($\log f_{\rm 2-10~keV}/f_{R} > +1$, i.e., optically-faint), 
the AGNs showing no significant broad H$\beta$ emission line, and 
the AGNs with a large X-ray absorption 
($N_{\rm H}$ larger than $10^{22}$  cm$^{-2}$), respectively. 
The X-ray luminosity in the 2--10 keV band of them ranges from 
$\sim 10^{43}$ erg s$^{-1}$ to $\sim 10^{46}$ erg s$^{-1}$, 
and the $N_{\rm H}$ ranges from $\sim 0 $ to $ \sim 10^{23}$ cm$^{-2}$.
The hard X-ray luminosities of the sample are one to two orders of
magnitudes larger than  those of AGN samples obtained with
{\it Chandra} and {\it XMM-Newton} at $z\lesssim1$. 
Therefore, the {\it ASCA} sample provides a good sample
to study the high-luminosity absorbed AGN population in the intermediate
redshift universe. 

\section{NIR and Optical Photometric Data Sources}

For the {\it ASCA} AGN sample, we made a cross-correlation with
2MASS Point Source Catalog (the second incremental data release).
The 10 $\sigma$ detection limits for a point source are 
15.8, 15.1 and 14.3 mag at $J$, $H$, and $K_{S}$ band, 
respectively in the catalog. 
We searched a 2MASS source each for an optical counterpart of the
{\it ASCA} AGN within a radius of 1$^{\prime\prime}$.2, 
except for one diffuse object. 
We found NIR counterparts for about half of the sample 
in the 2MASS catalog. 

In order to obtain NIR photometric data for 
the {\it ASCA} AGNs without 2MASS counterpart, 
$J$, $H$, and $K_{S}$ imaging observations were made
with Simultaneous Quad Infrared Imaging Device (SQIID) 
attached to the Kitt Peak National Observatory (KPNO) 2.1 m telescope 
during 4 nights from 2002 March 29 to April 1.
The SQIID can produce images of the same field in 
$J$, $H$, $K_{S}$, and $L$ bands simultaneously, with four 
$512 \times 512$ ALADDIN InSb arrays.
One pixel corresponds to 0.$^{\prime\prime}$69, 
giving a field of view of $\sim 5 \times 5$ arcmin$^2$. 
About 50 {\it ASCA} AGNs which did not have 2MASS counterparts 
at the time of the observing run were observed.
We took 5 images of each target with small ($\sim 20^{\prime\prime}$) offsets. 
A total integration time was 300 sec to 3600 sec for each object
depending on its faintness.
A typical seeing size (FWHM) during the observations was
$2^{\prime\prime}$ and  the sky conditions were photometric.

The data were reduced using IRAF in the following manner. 
After correcting bad pixels by interpolating a pixel value
from surrounding pixels, a dark frame was subtracted.
Flat fielding was done with flat frames which were constructed 
by averaging the normalized object frames taken within a
few hours in each band. 
A background sky was determined by a one-dimensional polynomial fitting
to the flat-fielded frame and was removed.
Using positions of bright stars (but not saturated) in a frame,
five frames for each target with small offsets were combined to one
frame by taking average. 

Since the SQIID adopted the filter system very close to
that of the 2MASS, a calibration of magnitude was made by using
objects cataloged in 2MASS Point Source Catalog and 
not cataloged in 2MASS Extended Source Catalog 
(i.e., excluding extended sources in the point source catalog). 
About 100 stellar-objects in 23 target fields were used. 
We restricted the magnitude ranges of the objects 
for the calibration to avoid non-linearity and poor S/N; 
12.5 mag $< J <$ 16.0 mag, 12.5 mag $< H <$ 15.5 mag, 
12.5 mag $< K_{S} <$ 15.5 mag. 
By taking account of colors of the objects and 
air mass of each frame, 
we determined a zero-point in each band
with 3$\sigma$ clipping algorithm.
The obtained zero-points are accurate with an uncertainty of 
$\pm 0.01$ mag for all the bands.

Photometry of the target objects was made as follows:
First, we estimated a residual local-sky value with 
the median of pixel values in an annular region around the object
(a 6 -- 15 pixel radius with a 6 -- 15 pixel width). 
Counts at the saturation radius of the growth curve of 
each target was adopted to be  total counts. 
The photometric error for each object was estimated by measuring
counts in an aperture with the same size for the
target object randomly putting into the object frame.
We put 10 apertures one time and repeated it 10 times for
each object frame.
Thus total 100 values were obtained for each target, 
and we assigned the rms value as the error of the target magnitude.
Resulting magnitudes and errors are listed in Table 1. 

About one year later after our observations, the 2MASS All-Sky Data 
Release was released.
Among the targets observed with the KPNO 2.1 m, 
which were not cataloged in the 2nd Incremental Data Release, 
23 sources were found in the All-Sky Catalog. 
The colors agree with each other, 
although a systematic error of $\sim$ 0.1 mag may exist. 
Since the KPNO data have higher S/N than those of 2MASS,
we adopt the KPNO data in this paper. 
The 2MASS All-Sky Data also provided us several new photometric data
for our sample, of which NIR data had not been available.

Additional observations were made on 2003 June 14 and 15 
with QUick InfraRed Camera (QUIRC) employing a 1024 $\times$ 1024
HgCdTe array attached to the University if Hawaii (UH) 88 Inch telescope;
9 of the remaining AGNs in the sample, 
for which no NIR photometric data were available or
only poor-S/N data were obtained, were observed 
in $J$, $H$, and $K^{\prime}$ bands.
The pixel scale of the camera was 0.$^{\prime\prime}$19, 
giving a field of view of 3.2 $\times$ 3.2 arcmin$^{2}$ 
and the seeing size during the observations was typically 
$\sim 0.^{\prime\prime}8$. 
We took more than 4 images of each target in each
band with small ($\sim 10^{\prime\prime}$) offsets. 
Total exposure time for each target ranged 5 min to 72 min 
depending on its faintness. 
During the observations, 4 UKIRT faint standard stars were 
observed for photometric calibration. 
All the data were reduced using IRAF in the same manner. 
Dark-frame subtraction, flat-fielding with self-flat frames, 
background subtraction, matching the seeings, 
and image stacking were performed. 
Errors on magnitudes of the targets were $\sim 0.1$ mag
or less in all the bands. 
$K^{\prime}$ magnitudes were converted to $K_{S}$ magnitudes 
by using the equations given by Wainscoat and Cowie (1992), 
UKIRT web site \footnote{http://www.jach.hawaii.edu/JACpublic/UKIRT/astronomy/calib/phot\_trans.html}, and 
Section VI.4.b of the 2MASS Explanatory Supplement \footnote{http://www.ipac.caltech.edu/2mass/releases/allsky/doc/sec6\_4b.html}. 
In summary, we obtained NIR photometric data for 104 out of
105 of the {\it ASCA} AGNs. 
The results are presented in Table 1. 
(One object without NIR data in our sample 
is 1AXG J035008$-$1149 (SE37).)

Optical photometric data are mainly taken from the APM catalog
(Irvin, Maddox, \& McMahon 1994). 
The $O$ and $E$  magnitudes of APM catalog are converted 
to $B$ and $R$ magnitudes by using the equations by Evans (1989). 
$B$ and $R$ magnitudes are obtained for $\sim 80\%$ and 
100\% of the sample objects, respectively, 
as shown in Table 1. 
Errors of $B$ and $R$ magnitudes are not available and 
we adopt 0.5 mag as a typical error. 
Supplemental optical magnitudes are taken from our own data 
(Akiyama et al. 2000a, 2003); they are noted in Table 1. 

\section{Results and Discussions}
\subsection{$K_S$ magnitudes and Hard X-ray Flux}
In Figure \ref{k_fx}, $K_{S}$ magnitude against 
2--10 keV flux is shown for our sample with open circles. 
Over-plotted filled circles, crosses, and open squares 
are the same as those in Figure \ref{L_z}. 
A symbol size of an open circle represents the 2--10 keV luminosity; 
large, middle, and small symbols indicate the AGNs with 
the 2--10 keV luminosity larger than $10^{45}$ erg s$^{-1}$, 
$10^{44-45}$ erg s$^{-1}$, and smaller than $10^{44}$ erg s$^{-1}$,
respectively. 
Dashed lines represent the constant X-ray to NIR flux ratio 
of $\log f_{\rm 2-10~keV}/f_{K_{S}}$ = 
$+$3, $+$2, $+$1, 0, $-$1, $-$2, and $-$3, from top to bottom. 
Most of the {\it ASCA} AGNs distribute between 
$\log f_{\rm 2-10~keV}/f_{K_{S}}$ = 0 and +1,
which is consistent with that of the mean spectrum of 
radio-quiet QSOs by Elvis et al. (1994) 
($\log f_{\rm 2-10~keV}/f_{K_{S}} \sim +0.4$). 
All but one of the optically-faint AGNs in the sample 
(shown with filled circles) fall above the line of 
$\log f_{\rm 2-10~keV}/f_{K_{S}}$ = +1. 
It is also notable that AGNs without broad H$\beta$ emission
lines (crosses), most of which are type 2 Seyferts 
at relatively lower redshift, 
tend to have relatively small X-ray to $K_S$ flux ratios. 
If the AGN components of them have 
the same X-ray to $K_S$ flux ratios as those of 
the other AGNs in our sample, 
the excess in $K_S$ flux at a given X-ray flux is 
considered to be caused 
by the contribution of their host galaxy light. 

\subsection{Color Distribution of the Hard X-ray Selected Sample}
The $J-K_{S}$ colors of the {\it ASCA} AGNs are plotted 
as a function of redshift in Figure \ref{jk_z}. 
Over-plotted filled circles, crosses, and open squares are 
the same as those in Figure \ref{L_z}. 
Small dots represent the optically-selected QSOs; 
counterparts in 2MASS (Second Incremental Data Release) of 
QSOs in Veron-Cetty \& Veron catalog (2000) excluding radio-loud QSOs 
(Barkhouse \& Hall 2001). 
Solid, dotted, and dashed lines show color tracks for 
the mean-QSO spectrum by Elvis et al. (1994), 
for that with an absorption of $A_V=1$ mag, 
and for a template spectrum of an Sbc galaxy by Coleman, Wu, \&
Weedman (1980) to represent the color of 
a host galaxy of an AGN, respectively. 
No evolution is considered. 
$B-K_{S}$ color against redshift for the {\it ASCA} sample 
is plotted in Figure \ref{bk_z}. 
Marks and tracks are the same as those in Figure \ref{jk_z}. 

In Figures \ref{jk_z} and \ref{bk_z}, 
the {\it ASCA} AGNs distribute between 
the Sbc-galaxy and the mean-QSO color tracks 
or around them at $z \lesssim 0.3$, 
while at $z \gtrsim 0.3$, they locate around the mean-QSO
color track. 
At the low redshift, most of the sample are low 
luminous AGNs (Seyfert class); the total colors are mixture of
those of host galaxies and nuclei. 
The objects at the high redshift are mostly luminous AGNs, 
resulting in that their colors show those of ordinary QSOs. 

Since almost all of the optically-selected QSOs have $J-K_{S}<2.0$ mag, 
we adopt $J-K_{S}>2.0$ mag as a criterion for red AGNs, which 
is the same as that adopted by Cutri et al. (2001). 
In our sample, two out of 104 AGNs show red color. 
Therefore, the fraction of the red AGNs 
in the {\it ASCA} sample is $2\pm1\%$. 
If the error bars of $J-K_{S}$ color for the sample AGNs are considered, 
the maximum of the fraction is $\sim 8\pm3\%$. 
Thus among the AGNs with  the 2--10 keV flux larger than 
$\sim 10^{-13}$ erg s$^{-1}$ cm$^{-2}$, 
the fraction of red AGNs is not 
significantly larger than that in the optically-selected sample 
($\sim 5\%$) (Barkhouse \& Hall 2001). 

On the other hand, the $B-K_{S}$ color distribution of 
the {\it ASCA} AGNs seems to be biased toward red 
compared with the optically-selected QSOs (dots) 
particularly at $z\lesssim0.3$. 
Most of the {\it ASCA} AGNs showing red $B-K_{S}$ color 
are type 2 Seyferts, so the origin of their red color 
is likely to be host-galaxy light. 

\subsection{Individual Notes on Red AGNs in the {\it ASCA} Sample}
There are only two AGNs with $J-K_{S}$ colors redder than 2 mag in
our sample;
1AXG J121854$+$2957 (NO07) with $J-K_{S} = 2.98$ mag and
1AXG J172938$+$5230 (NO10) with $J-K_{S} = 2.03$ mag.
There are six AGNs at $z>0.6$ showing similarly
red ($J-K_{S} \sim 2$ mag) colors, which are redder than
those of the other AGNs in the redshift range. 
These 8 red AGNs are marked with 
over-plotted pluses in Figure \ref{k_fx}. 
We discuss these objects and the origin of the red colors 
in this section. 
In each panel of Figure \ref{indiv}, 
the NIR and optical photometric data points 
of each red AGN are shown by filled and open squares 
with the error, respectively. 
Solid and dashed lines 
indicate the mean-QSO spectrum (Elvis et al. 1994) with the absorption 
shown in the figure (Galactic extinction curve by 
Cardelli, Clayton, \& Mathis (1989) is used) and 
the Sbc-galaxy template (Coleman et al. 1980) without absorption, 
respectively. 
These model spectra 
are independently fitted to the NIR data points of each AGN 
with the least-square method. 
It should be noted that the optical and NIR data are not coeval 
so that the variability may affect the spectral energy distribution 
(SED) for each AGN. 

\subsubsection{{\it ASCA} AGNs with $J-K_{S}>2.0$ mag}
\begin{figure}
\plotone{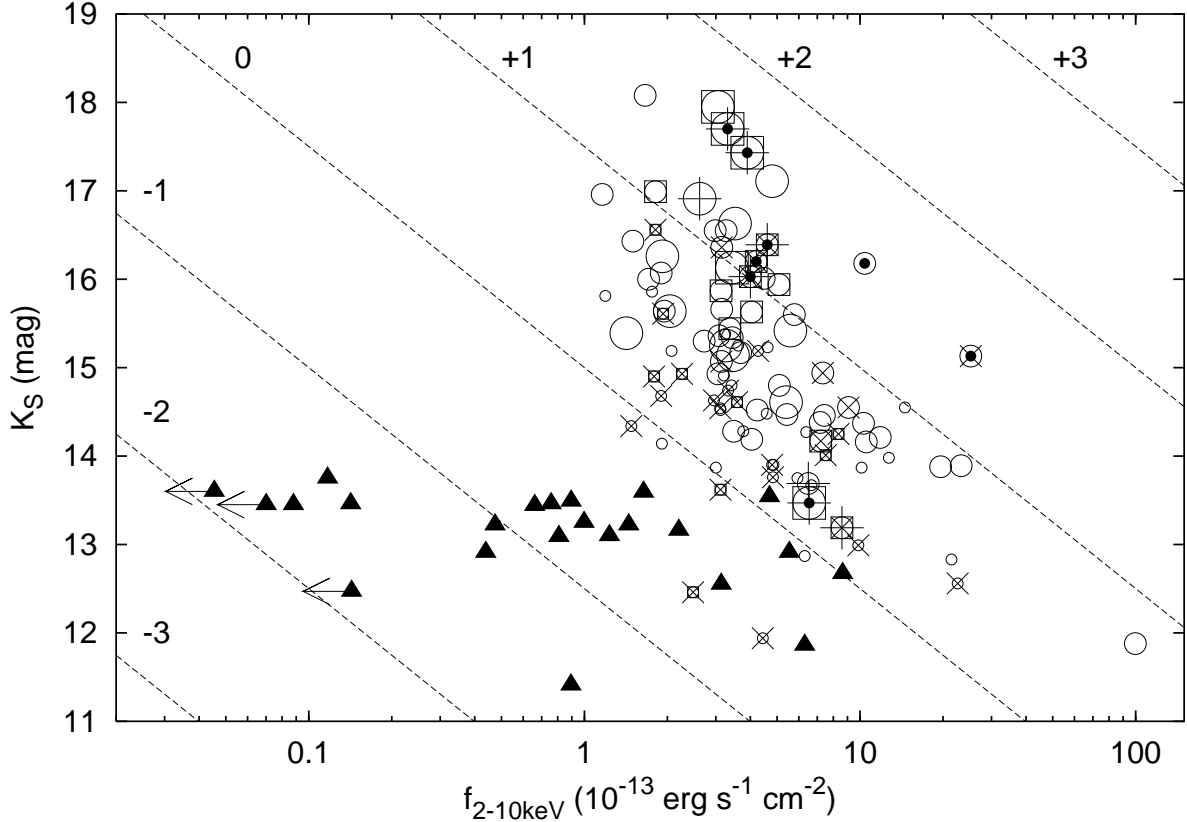}
\caption{
$K_{S}$ magnitude against 2--10 keV X-ray flux 
for {\it ASCA} AGNs (open circles). 
Symbol sizes represent the absorption corrected rest-frame
luminosity in the 2--10 keV band;
large, middle, and small symbols represent 
AGNs with $L_{\rm 2-10~keV}>10^{45}$ erg s$^{-1}$, 
$10^{44-45}$ erg s$^{-1}$, and $<10^{44}$ erg s$^{-1}$, respectively. 
Over-plotted filled circles, crosses, and open squares 
are the same as those in Figure \ref{L_z}. 
Over-plotted pluses indicate the red AGNs; 
two AGNs with $J-K_S>2$ mag and six AGNs with $J-K_S\sim2$ mag at $z>0.6$, 
which are described in Section 4.3. 
Filled triangles show the 2MASS red ($J-K_S>2$ mag) AGNs 
(Smith et al. 2002, Wilkes et al. 2002). 
Dashed lines represent the constant X-ray to NIR flux ratio 
of $\log f_{\rm 2-10~keV}/f_{K_{S}}$ = $+$3,$+$2,$+$1,0,$-$1,$-$2, 
and $-$3 from top to bottom. 
}
\label{k_fx}
\end{figure}

\begin{figure}
\epsscale{1.0}
\plotone{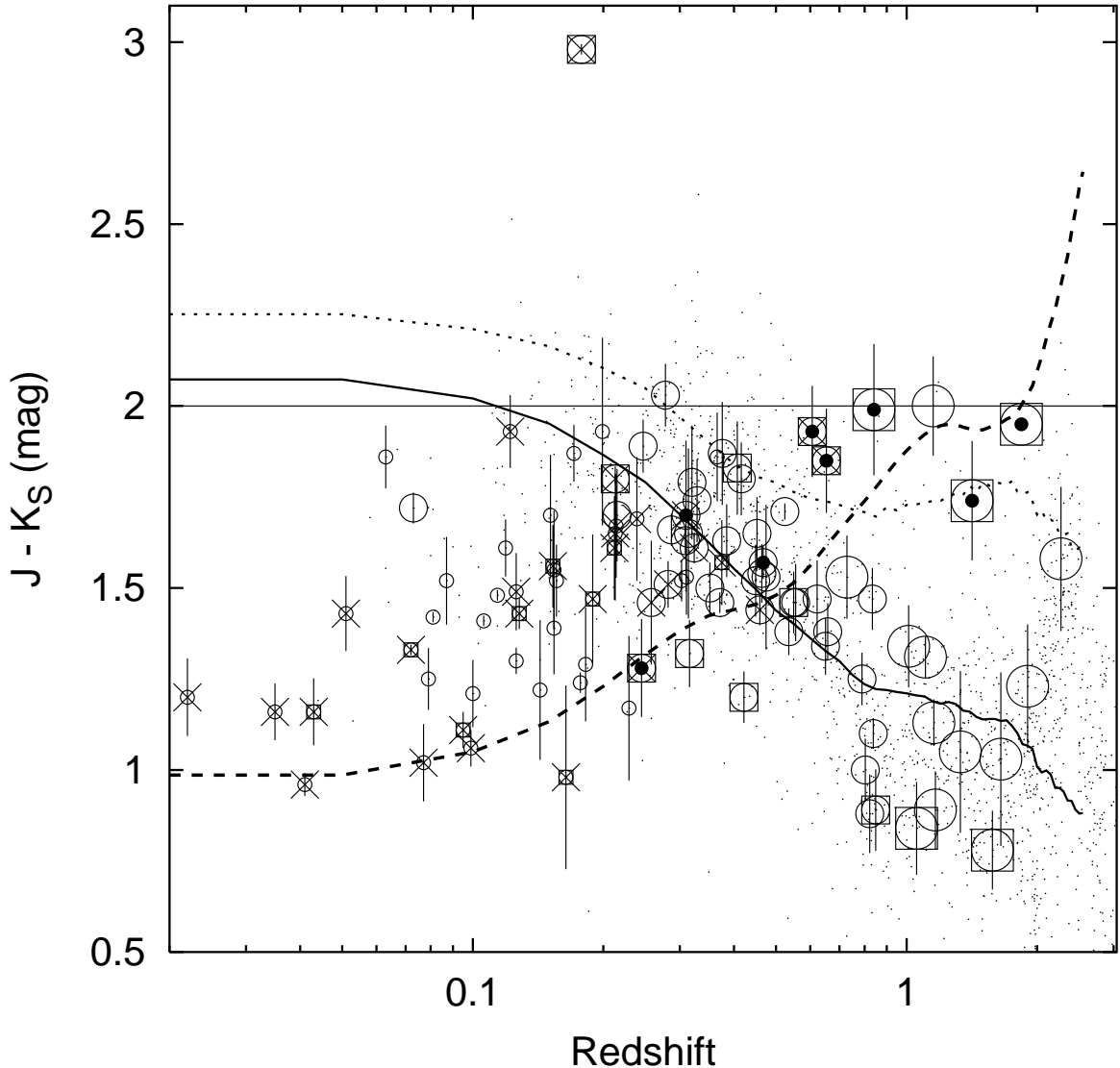}
\caption{
$J-K_{S}$ color vs redshift for the {\it ASCA} AGNs
(open circles). 
Over-plotted filled circles, crosses, and open squares 
are the same as those in Figure \ref{L_z}. 
Small dots 
indicate color distribution of optically-selected QSOs measured
in the 2MASS survey (Barkhouse \& Hall 2001).
$J-K_{S}$ color track for the mean-QSO spectrum by Elvis et al. (1994), 
that for the same mean-QSO spectrum with absorption with $A_V=1$ mag, 
and that for an Sbc galaxy taken from Coleman et al. (1980) 
are shown with solid, dotted, and dashed lines, respectively.
No spectral evolution is considered.
}
\label{jk_z}
\end{figure}

\begin{figure}
\plotone{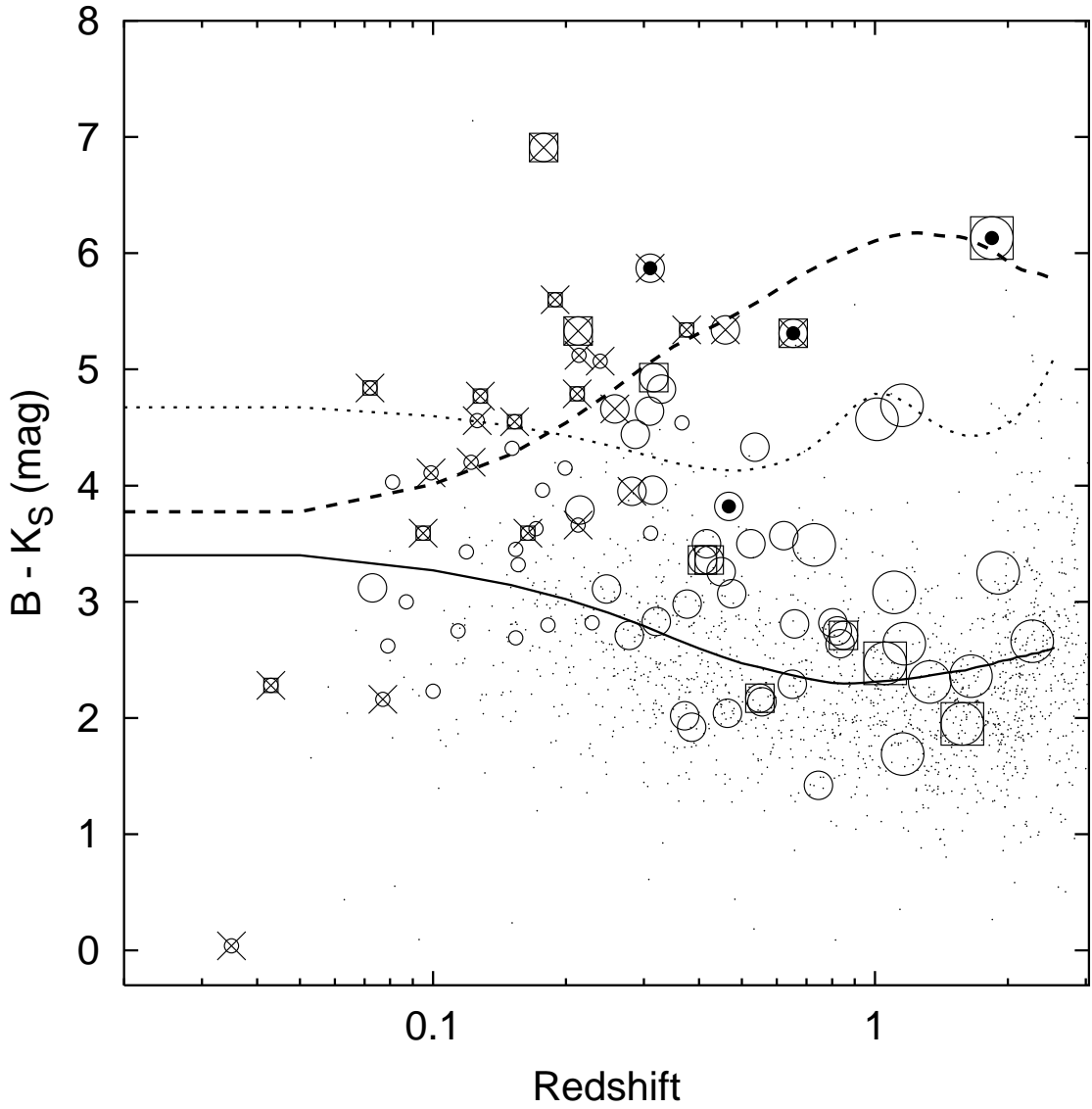}
\caption{
Same as figure \ref{jk_z}, but for $B-K_{S}$ color.
}
\label{bk_z}
\end{figure}

\begin{figure}
 \begin{center}
  \begin{tabular}{ccc}
   \resizebox{65mm}{!}{\includegraphics{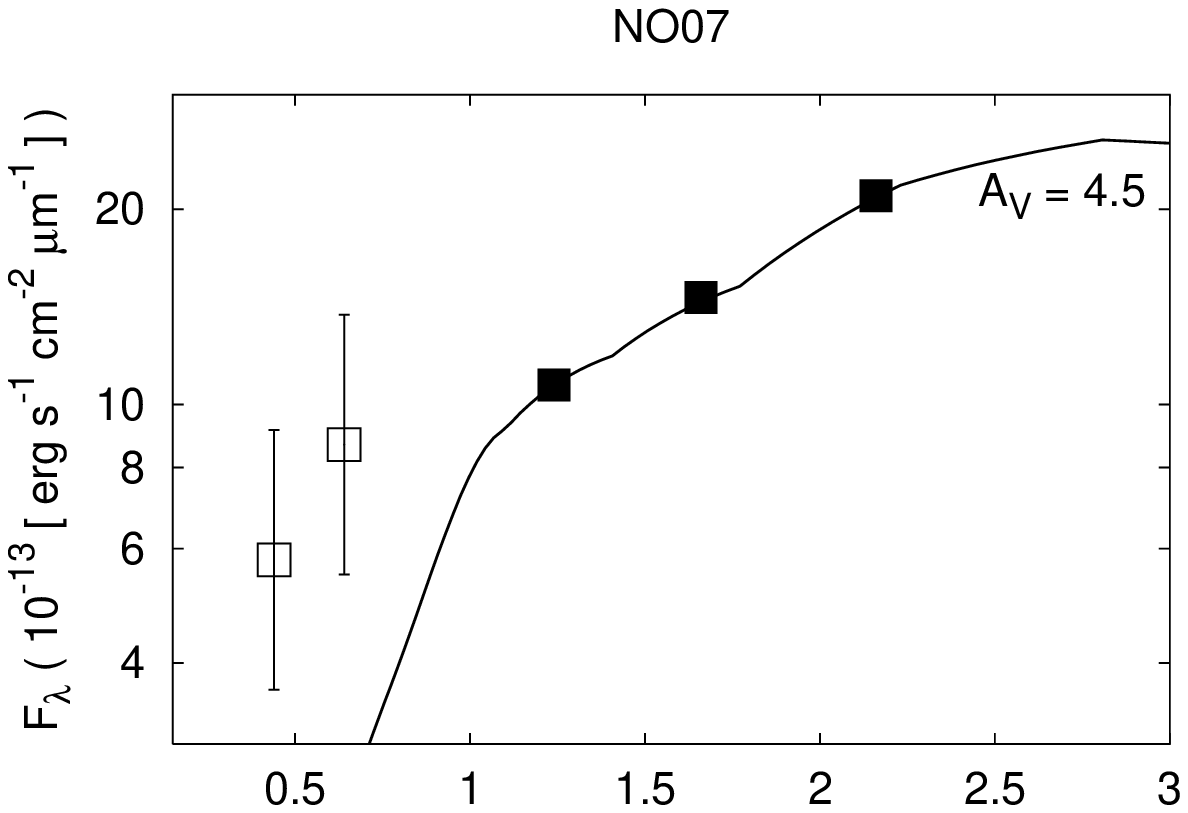}} &
   \resizebox{65mm}{!}{\includegraphics{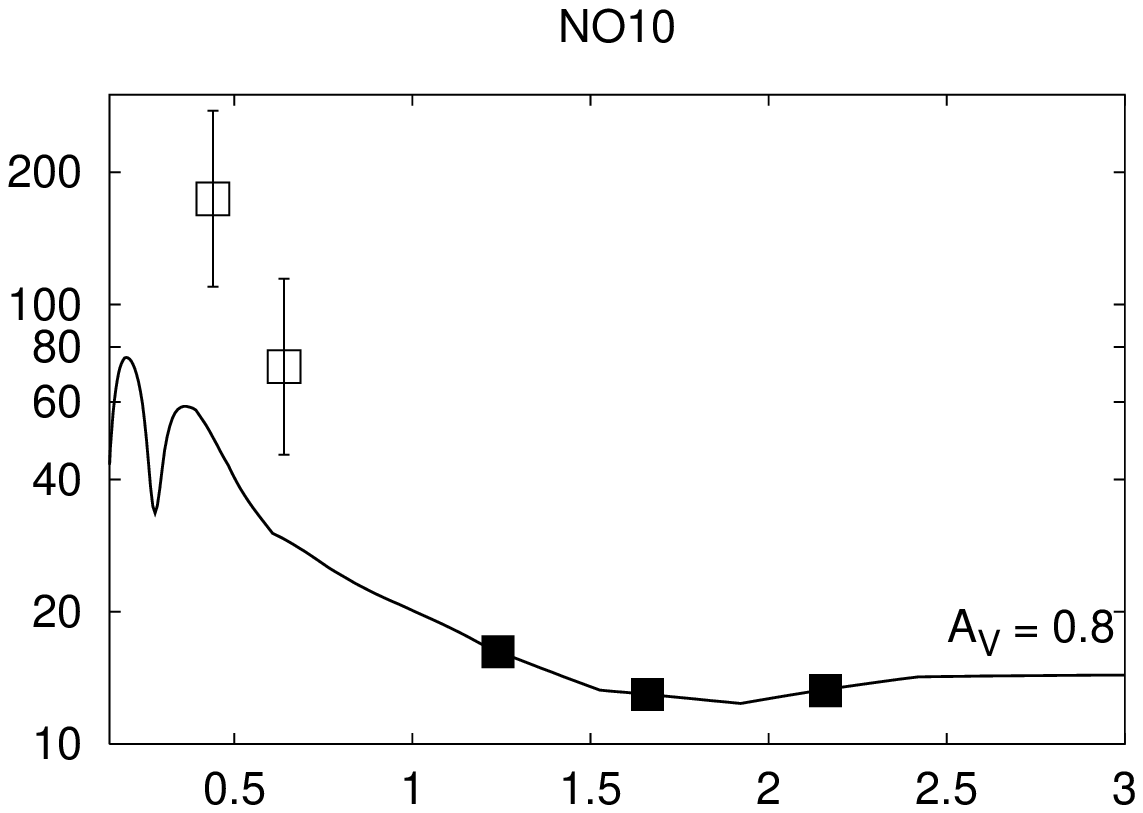}} \\
   \resizebox{65mm}{!}{\includegraphics{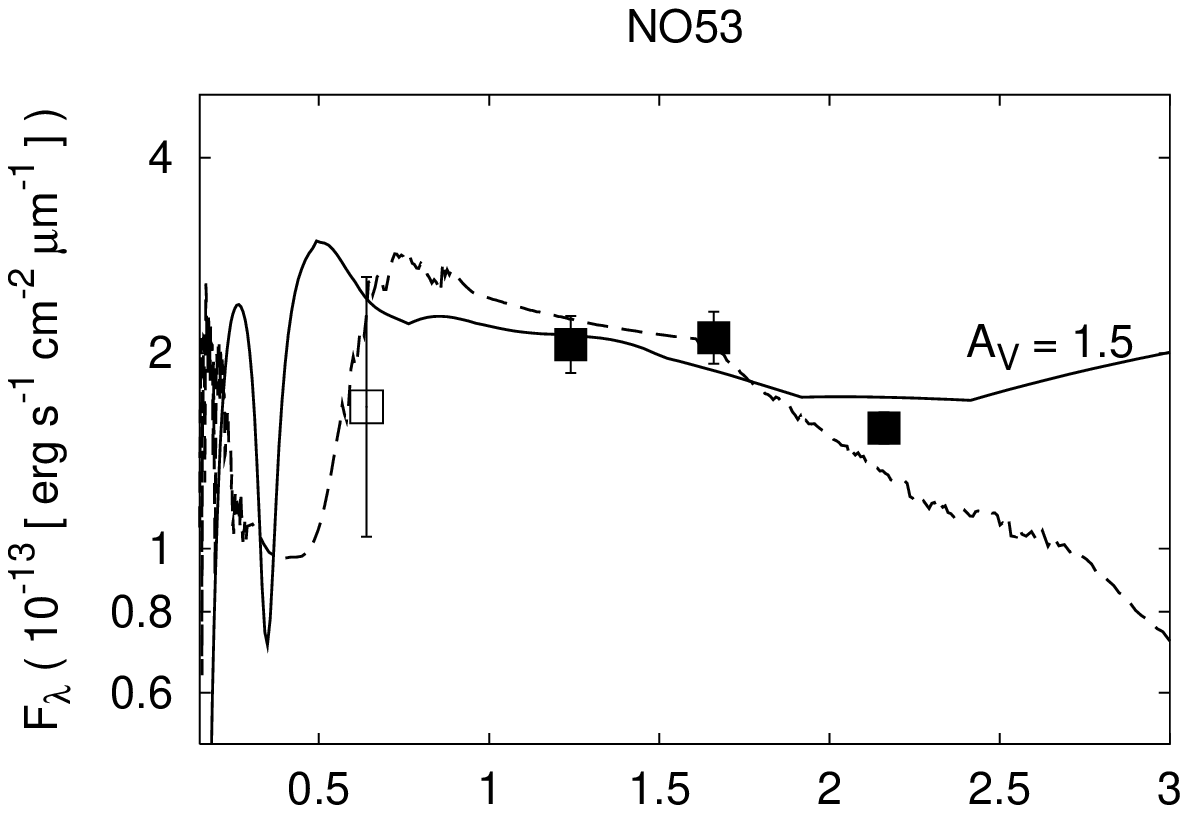}} &
   \resizebox{65mm}{!}{\includegraphics{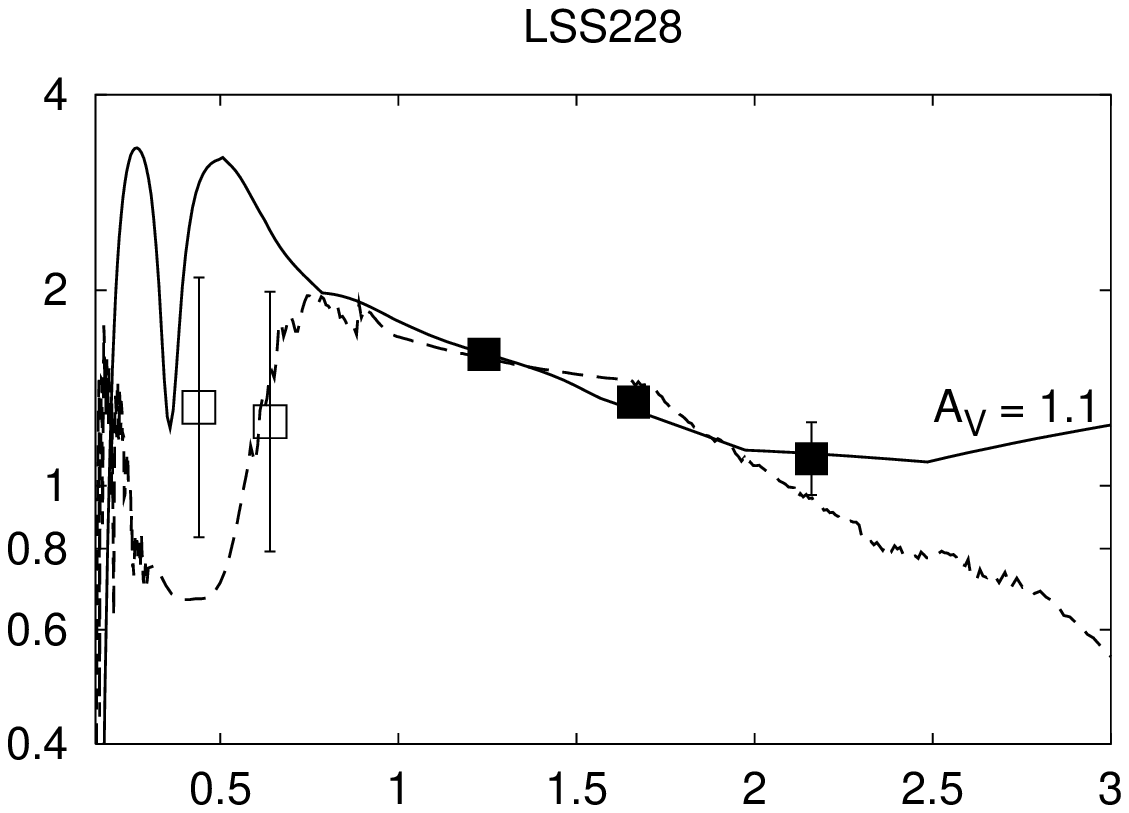}} \\
   \resizebox{65mm}{!}{\includegraphics{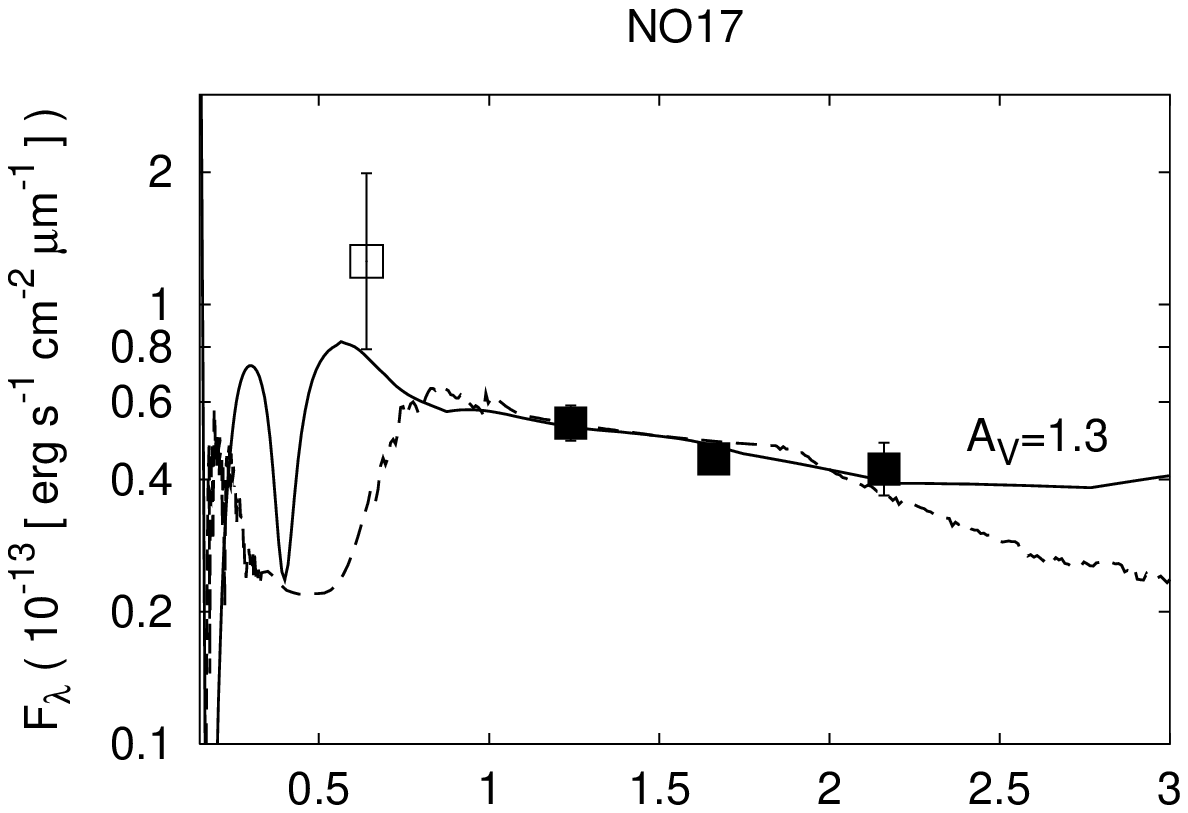}} &
   \resizebox{65mm}{!}{\includegraphics{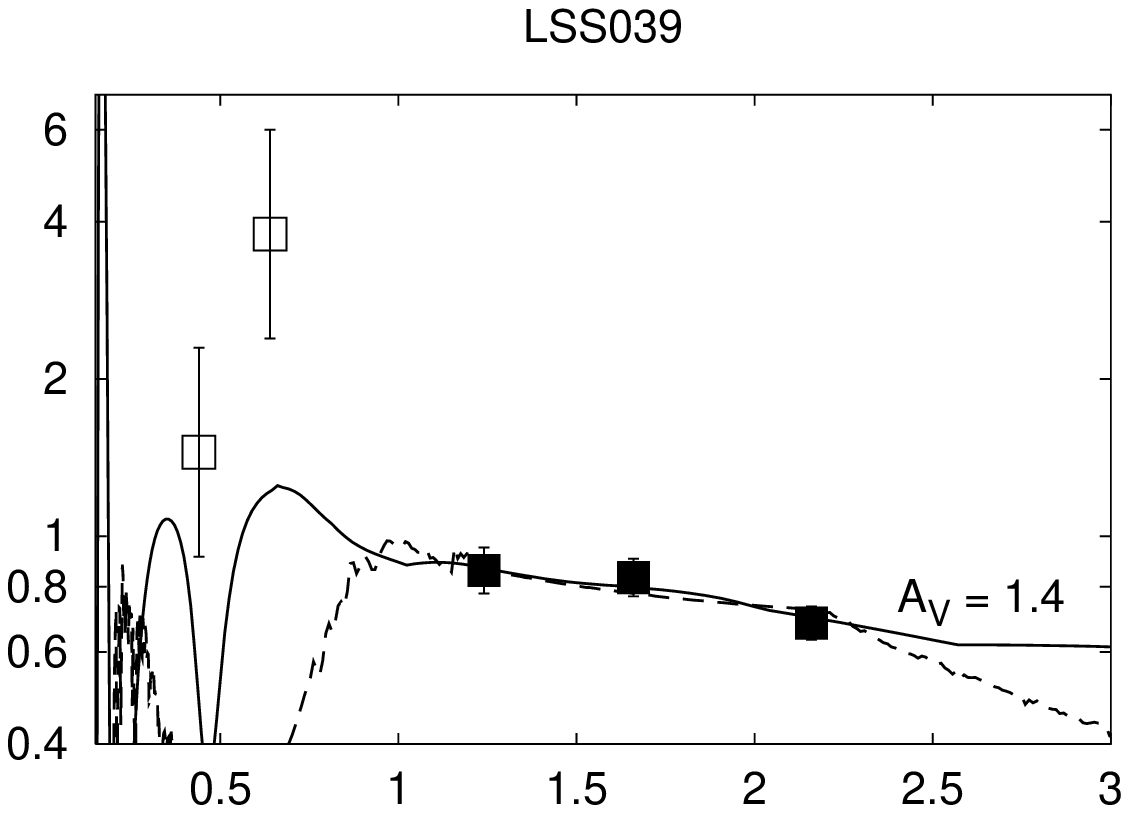}} \\
   \resizebox{65mm}{!}{\includegraphics{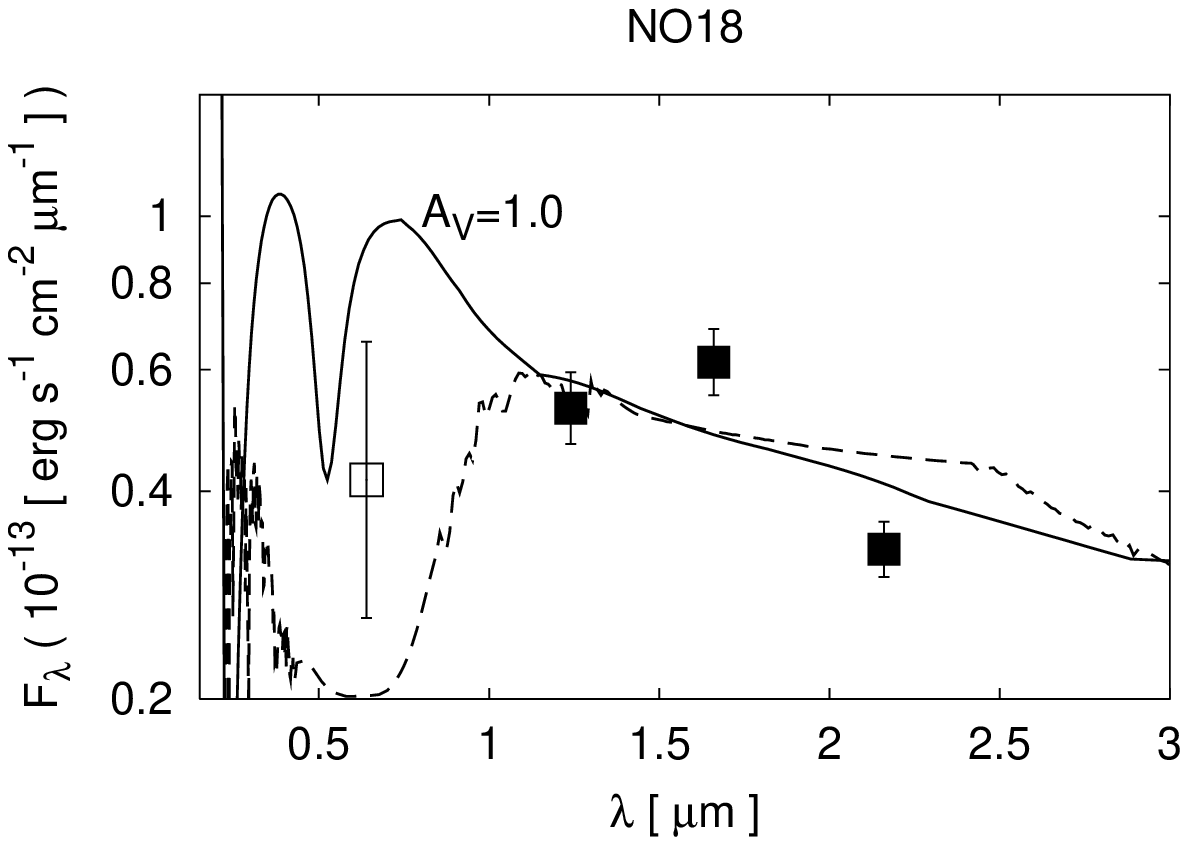}} &
   \resizebox{65mm}{!}{\includegraphics{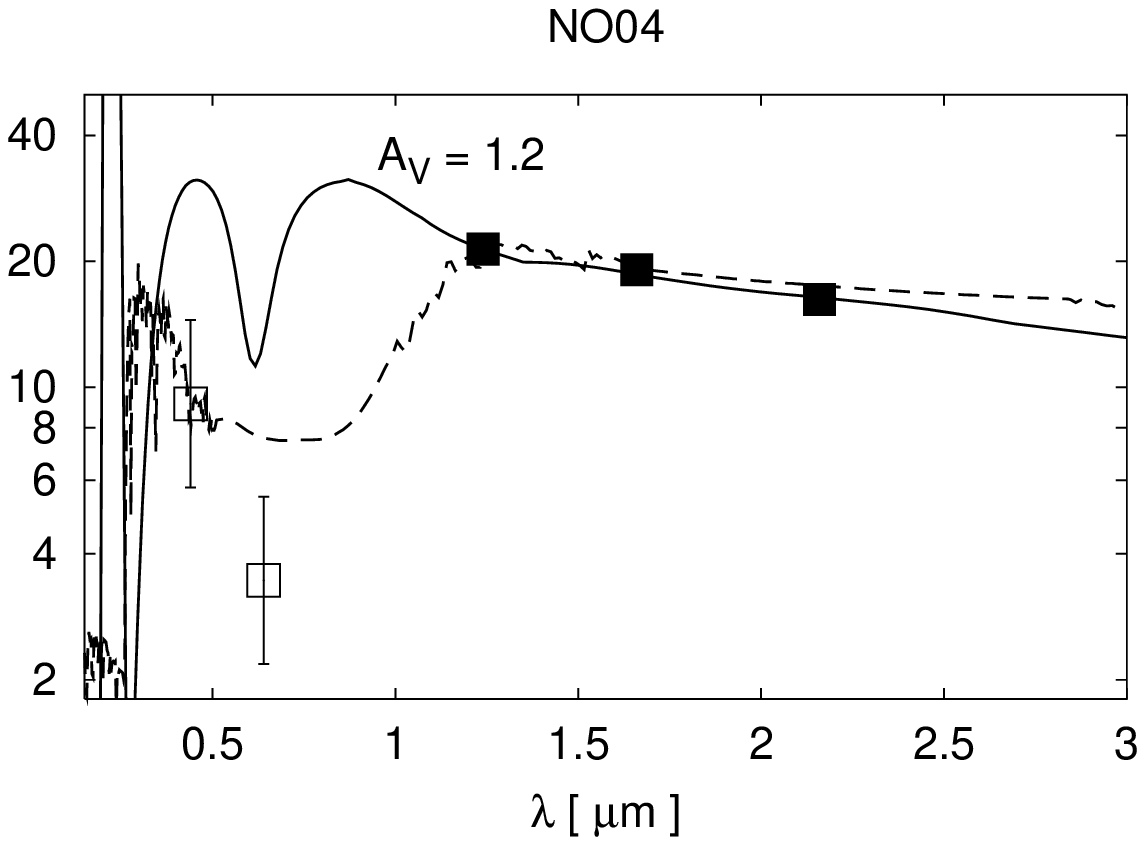}} \\
  \end{tabular}
 \end{center}
\caption{
Spectral energy distributions of the red AGNs in the sample; 
two AGNs with $J-K_S>2$ mag (NO07, NO10) and six AGNs with 
$J-K_S\sim2$ mag at $z>0.6$ (NO53, LSS228, NO17, LSS039, NO18, NO04), 
which are described in Section 4.3. 
Filled and open squares show the photometric data of 
NIR ($J$, $H$, $K_{S}$) and optical ($B$, $R$) bands, respectively. 
Solid lines indicate the best-fit reddened QSO spectrum 
(in detail see section 4.3) with $A_V$ shown in the panels.
Dashed lines show the best-fit Sbc-galaxy template spectrum 
by Coleman et al.(1980) without absorption. 
Errors on $B$ and $R$ magnitudes are taken to be 0.5 mag.}
\label{indiv}
\end{figure}

1AXG J121854$+$2957 (NO07) has the reddest $J-K_{S}$ color (2.98 mag) 
among the sample. 
It is an AGN at $z = 0.178$ with 
an intrinsic (i.e., absorption corrected) X-ray luminosity of 
$L_{\rm 2-10~keV} = 1.1 \times 10^{44}$ erg s$^{-1}$. 
The $H\beta$ emission line of this AGN is narrow, but the $H\alpha$ 
emission line has a faint wing (Fiore et al. 1999). 
Thus this object seems to be a Seyfert 1.9. 
The SED of the AGN from $B$ to $K_{S}$ is shown in Figure \ref{indiv}. 
The mean-QSO spectrum with $A_{V} = 4.5$ mag 
can well reproduce the SED from $J$ to $K_{S}$. 
The $A_{V}$ is considerably smaller than that expected 
from the X-ray absorption of this AGN; 
by using the Galactic dust-to-gas ratio of 
$A_V/N_{\rm H} = 5.6 \times 10^{-22}$ mag cm$^2$ 
(Predehl \& Schmitt 1995), 
$A_{V} = 46_{-4}^{+6}$ mag is estimated from the 
$N_{\rm H} = 8.2_{-0.7}^{+1.1} \times 10^{22}$ cm$^{-2}$ 
derived from the X-ray spectrum of this AGN obtained with 
{\it XMM-Newton} (Loaring, Page, \& Ramsay 2003). 
If the mean-QSO spectrum with $A_{V} = 4.5$ mag is assumed, 
$B$ and $R$ -band data show excesses; 
a contribution from the galaxy 
may brighten $B$ and $R$ -band magnitudes. 
It is consistent with a fact that 
this AGN appears to be nearly point-like in NIR images 
but extended in optical images. 
Although the NIR SED can be also reproduced by a galaxy spectrum 
with an absorption of $A_V = 8.4$ mag and with a luminosity of about 17 $L^*$, 
it is unlikely because the host galaxy seems to be too bright. 
In Figure \ref{k_fx}, this AGN locates near the line of 
$\log f_{\rm 2-10~keV}/f_{K_{S}}=0$, which is close to a location 
occupied by a part of the 2MASS red AGNs shown with 
filled triangles. 

1AXG J172938$+$5230 (NO10) is a broad-line AGN at $z = 0.278$ with 
an intrinsic X-ray luminosity of 
$L_{\rm 2-10~keV} = 1.7 \times 10^{44}$ erg s$^{-1}$. 
It looks like a point source on the NIR and optical images. 
Its SED from $J$ to $K_S$ is well reproduced with the mean-QSO
spectrum with $A_V = 0.8$ mag, but in this case, 
$B$ and $R$ -band magnitudes are too bright. 
The mean-QSO spectrum without absorption 
can reproduce the optical and NIR SED reasonably well. 
It is consistent with the fact that the
deviation of $J-K_{S}$ color from the mean-QSO track is small
as seen in Figure \ref{jk_z}. 
In Figure \ref{k_fx}, this AGN also locates close to the line of 
$\log f_{\rm 2-10~keV}/f_{K_{S}}=0$. 

\subsubsection{{\it ASCA} AGNs with $J-K_{S} \sim 2.0$ mag at $z > 0.6$}
1AXG J160118$+$0844 (NO53) is an AGN at $z = 0.606$ with 
an intrinsic X-ray luminosity of 
$L_{\rm 2-10~keV} = 5.9 \times 10^{44}$ erg s$^{-1}$. 
Its X-ray luminosity and narrow $H\beta$ emission line suggest that 
it is a candidate of a type 2 QSO. 
It is an optical-faint, radio-loud (figure 15 of Akiyama et al.(2003)) AGN.
In the optical and NIR images, it appears to be a point source. 
The SED of this AGN is roughly reproduced with the mean-QSO spectrum with
an absorption of $A_{V} = 1.5$ mag, which is very much
smaller than that estimated from the X-ray absorption 
($N_{\rm H} = 3.0_{-1.2}^{+1.6} \times 10^{22}$ cm$^{-2}$ 
corresponding to $A_{V} = 17_{-7}^{+9}$ mag). 
The SED can also be reproduced with a galaxy spectrum
with an intrinsic luminosity of about 6 $L^*$.

AX J131831$+$3341 (LSS228) is an AGN at $z = 0.653$ with 
an  intrinsic X-ray luminosity of $L_{\rm 2-10~keV} = 7.8 \times 10^{44}$ 
erg s$^{-1}$. 
Although its $H\beta$ emission line is narrow, 
it shows a broad emission line of Mg~II$\lambda$2800.
The absorption derived from X-ray spectrum is
 $N_{\rm H} = 1.1_{-0.4}^{+0.4} \times 10^{22}$ cm$^{-2}$.
This object is an optical-faint object.
The SED of this object is roughly consistent with the mean-QSO
spectrum with $A_{V} = 1.1$ mag or the galaxy template 
with a luminosity of about 5 $L^*$. 
However, its optical and NIR images show that it has a nuclear
component and an extended component, and we need to discuss the SED
 for each component. 
Akiyama et al. (2000b) and Akiyama and Ohta (2001) discuss this in detail. 
Its nucleus is very red; $I - K = 4.29$ mag, which corresponds to 
the absorption of $A_{V} \sim 3$ mag for the power-law continuum
of an index of $\alpha = -0.5$. 
The NIR and optical colors of the extended component are consistent 
with the post-starburst nature of the host galaxy. 

1AXG J210738-0512 (NO17) is a broad-line AGN at $z = 0.841$ with 
an intrinsic X-ray luminosity of $
L_{\rm 2-10~keV} = 1.2 \times 10^{45}$ erg s$^{-1}$. 
It appears nearly a point source in the NIR image but a 
slight elongation may be seen in the optical image.
This is an optical-faint, radio-loud object.
Its NIR SED is consistent with the absorbed mean-QSO spectrum 
with $A_{V} = 1.3$ mag, which is very much smaller than that
inferred from the X-ray absorption 
($N_{\rm H} = 4.3_{-1.7}^{+2.1} \times 10^{22}$ cm$^{-2}$
corresponding to $A_V = 24_{-10}^{+12}$ mag).
The red NIR colors are also accountable with a galaxy 
with a luminosity of about 4 $L^*$. 
However we prefer the former interpretation, because 
it has a large X-ray luminosity, shows  the broad emission line,
and has a point-like morphology. 

AX J131021+3019 (LSS039) is a broad-line AGN at $z=1.152$ \footnote{
The identification of this X-ray source was not correct in Akiyama
et al. (2000a).  Subsequent $XMM-Newton$ observation revealed that
object Z by Akiyama et al. (2000a) is the correct identification 
(See also Ueda et al. 2003).}
with an intrinsic X-ray luminosity of 
$L_{\rm 2-10~keV} = 1.8 \times 10^{45}$ erg s$^{-1}$. 
The object is point like in the optical image, but there is a hint
of the presence of an extended feature in the NIR images.
Its NIR SED is reasonably reproduced with the mean-QSO spectrum with 
$A_{V} = 1.4$ mag or the galaxy spectrum with a luminosity of 
about 17 $L^*$. 
Again we prefer the former interpretation with the same reasons
as those for NO17.

1AXG J233200+1945 (NO18) is an AGN at $z = 1.416$ with 
an intrinsic X-ray luminosity of 
$L_{\rm 2-10~keV} = 3.5 \times 10^{45}$ erg s$^{-1}$. 
There is a hint of a broad Mg~II emission line. 
This is an optical-faint source.
In the optical and NIR images, it appears to be a point source.
The SED of the AGN is consistent with the mean-QSO spectrum with 
an absorption of $A_{V} = 1.0$ mag, which is nearly two orders of magnitude
smaller than that estimated from the X-ray absorption 
($N_{\rm H} = 1.1_{-0.5}^{+0.6} \times 10^{23}$ cm$^{-2}$
corresponding to $A_V = 62_{-28}^{+34}$ mag). 
Although the galaxy spectrum with a luminosity of about 17 $L^*$ 
can also reproduce the NIR SED, 
we favor the former interpretation with the same reasons
as those for NO17.

1AXG J150423+1029 (NO04) is a broad-line AGN at $z = 1.839$ with 
an intrinsic X-ray luminosity of 
$L_{\rm 2-10~keV} = 1.2 \times 10^{46}$ erg s$^{-1}$. 
It is an optical-faint object. 
It is a radio-loud, bright quasar. 
In the optical and NIR images, it appears to be a point source.
Its SED is consistent with the mean-QSO spectrum with 
an absorption of $A_{V} = 1.2$ mag, which is smaller 
than that estimated from the X-ray absorption 
($N_{\rm H} = 5.4_{-1.7}^{+1.8} \times 10^{22}$ cm$^{-2}$
corresponding to $A_V = 30_{-10}^{+10}$ mag)  
by more than one order of magnitude. 
The SED is also consistent with 
the galaxy spectrum with a luminosity of about 1000 $L^*$,
which seems to be exceptionally bright. 
Therefore, we prefer the former interpretation. 

In summary, the relatively-red colors in NIR band of 
these six AGNs at $z>0.6$ 
are presumably originated from nuclear light with 
absorption of $A_V \sim 1-2$ mag or galaxy light 
with a luminosity of $\sim 4-17 L^*$. 
For many of these AGNs, we favor the former interpretation 
because of their point-like appearance, high luminosity in X-ray, 
and/or existence of a broad emission line. 5 out of 6 these AGNs 
have hard X-ray to $K_S$ band flux ratios larger than 
$\log f_{\rm 2-10 keV}/f_{K_S} = +1$. 
It suggests that these AGNs are different population 
from the 2MASS red AGNs, which have hard X-ray to 
$K_S$ band flux ratios of $\log f_{\rm 2-10 keV}/f_{K_S} \lesssim 0$. 

\subsection{Dust-to-Gas Ratios in QSOs}
\begin{figure}
\plotone{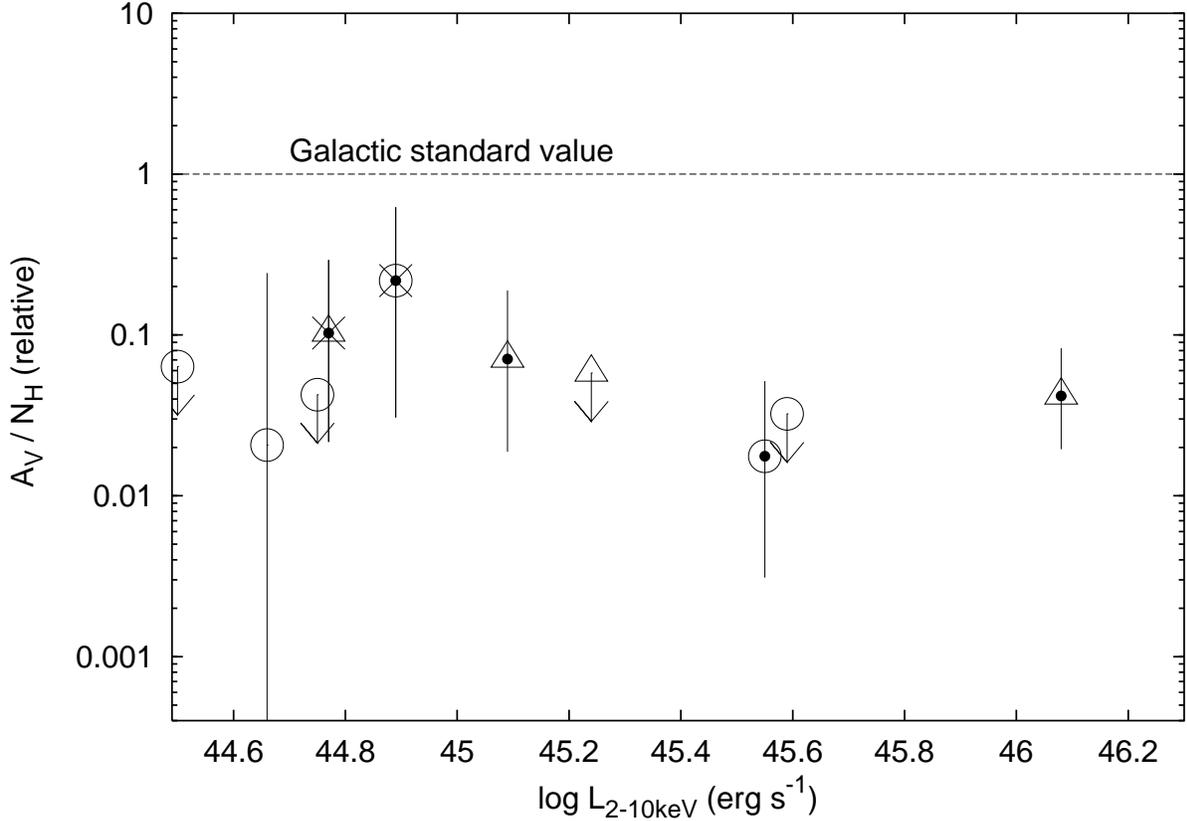}
\caption{
$A_{V}/N_{\rm H}$ value normalized by the Galactic value of 
$5.6 \times 10^{-22}$ mag cm$^2$ against absorption corrected X-ray
luminosity in the rest-frame 2--10 keV band for 
the sample objects with $N_{\rm H} \ge 10^{22}$ cm$^{-2}$
and $L_{\rm 2-10~keV} \ge 10^{44.5}$ erg s$^{-1}$.
Open circles and open triangles refer to the 
radio-quiet and radio-loud QSOs, respectively. 
Over-plotted filled circles and crosses 
are the same as in Figure \ref{L_z}. 
Arrows show upper limits of $A_{V}/N_{\rm H}$ value. 
The $N_{\rm H}$ is estimated from the X-ray spectrum with an assumption 
of an intrinsic power-law photon index of 1.9, 
and $A_{V}$ is calculated from the observed $J-K_{S}$ color by
assuming that the intrinsic spectrum of a QSO 
is the same as that of the mean-QSO spectrum by Elvis et al. (1994). 
}
\label{avnh_L}
\end{figure}

In this subsection, we estimate the dust-to-gas ratios 
of QSOs in the {\it ASCA} sample to 
investigate whether the anomalous dust-to-gas ratios found 
in low-luminous AGNs (Seyferts) 
are also seen in QSOs or not. 
In Figure \ref{avnh_L}, $A_{V}/N_{\rm H}$ values normalized to 
the Galactic one 
are plotted against 2--10 keV luminosity for the X-ray absorbed 
($N_{\rm H} \ge 10^{22}$ cm$^{-2}$) QSOs ($L_{\rm 2-10~keV} 
\ge 10^{44.5}$ erg s$^{-1}$) among the {\it ASCA} sample. 
Open circles and open triangles show the 
radio-quiet and radio-loud QSOs of the sample, respectively. 
Over-plotted crosses and filled circles are 
the same as those in Figure \ref{L_z}. 
The $N_{\rm H}$ is estimated from the X-ray spectrum 
as described in section 2, and the $A_{V}$ is
calculated from the observed $J-K_{S}$ color on the assumption that 
the intrinsic spectrum of a QSO is the same as 
that of the mean-QSO spectrum by Elvis et al. (1994).
Since both X-ray and $K_S$ luminosities are very high
(the absolute $K_S$ magnitudes of them are brighter than
$\sim -26$ mag which is 1.5 mag brighter than $M_{K}^{*}$ for local
galaxies (Huang et al. 2003)), 
the contribution to NIR
light from a host galaxy is assumed to be negligible. 
The uncertainty of $A_{V}/N_{\rm H}$ shown in the figure 
is calculated from the errors of $J-K_{S}$ color and of 
$N_{\rm H}$ for each QSO as well as the dispersion of distribution
in $J-K_{S}$ color of the optically-selected QSOs.
When no nominal optical absorption is obtained, 
an upper possible limit on $A_V/N_{\rm H}$ is calculated 
by considering the uncertainty, and is shown with an arrow. 

Figure \ref{avnh_L} shows that all the sample QSOs have 
$A_{V}/N_{\rm H}$ values smaller than the Galactic value by an order of 
magnitude or more\footnote{It should be noted that 
about half (4/10) of the sample QSOs are radio-loud. 
Since radio-loud QSOs tend to have flatter X-ray spectrum than 
radio-quiet ones, their $N_{\rm H}$ values 
derived by assuming an intrinsic power-law spectrum of
photon index of 1.9 may be over-estimated. 
However, even if an intrinsic photon index of 1.7 is assumed, 
their $N_{\rm H}$ values are reduced by only 20\% typically.}. 
This trend is the same as that seen in less luminous AGNs
by Maiolino et al. (2001a), and it is indicated that
the anomaly of the dust-to-gas ratio also occur in QSO-class AGNs.
Furthermore, all the objects in Figure \ref{avnh_L}
show the presence of broad components at least in one of
the emission lines of C~IV$\lambda$1549, Mg~II$\lambda$2800, 
H$\beta$, and H$\alpha$, which also indicates 
the lack of the heavy extinction in the optical/UV region.   
The results show not only the existence of a population that
has a type 1 nature in the optical/UV band but show a type 2 nature
in the X-ray band, but also show that it is not a rare population.
Since a fraction of this population in our sample is 
about 30\% among the sample with the X-ray luminosity larger than
$10^{44.5}$ erg s$^{-1}$, such fraction of optical/UV-selected
type 1 QSOs known to date may have large amount of dust-less 
circumnuclear gases. 

\subsection{Comparison with 2MASS red AGNs}
Cutri et al. (2001) revealed the presence of many red ($J-K_{S} >2$ mag)
AGNs at $z \lesssim$ 0.3, by making optical follow-up 
for the red AGNs candidates selected with 
the 2MASS catalog (2MASS red AGNs).
The optical follow-up spectroscopy of a part of them indicates that they are
mixture of type 1 and type 2 AGNs; 
the number of type 2 is about one third of that of type 1.
They claim that the number density of red type 1 AGNs 
is comparable to that of known QSOs 
selected in optical or UV. 

How many these red AGNs are predicted to be detected 
in our surveys with {\it ASCA} ?
The total survey area of ALSS and AMSSn is $\sim$ 75 deg$^2$. 
Since the surface number density of red AGNs is 
$\sim$ 0.2 deg$^{-2}$ for $K_S \lesssim 14$ mag (Cutri et al. 2001), 
the region of our surveys should contain about 15 red AGNs 
with $K_S \lesssim 14$ mag. 
Figure \ref{k_fx} (filled triangles) shows 
$K_S$ magnitude and X-ray flux in the 2--10 keV band for 
a well-defined subsample of the 2MASS red AGNs 
observed with {\it Chandra} (Wilkes et al. 2002). 
About 20\% of them 
have X-ray flux larger than the detection limit 
of AMSSn\footnote{Strictly speaking, since the fluxes of the 2MASS 
red AGNs are simply converted from the Chandra 0.3--8 keV
count rate assuming a photon index of 2, the true 2--10 keV fluxes 
could be significantly larger than the plotted values 
in Figure \ref{k_fx} if they are 
heavily absorbed as inferred from their hard spectra (Wilkes et al. 
2002). Assuming a typical absorption of 
$N_{\rm H} = 10^{22.5}$ cm$^{-2}$ at 
a redshift of 0.2, we expect that about 35\% of them have count 
rates larger than the detection limit of the AMSSn.}. 
Therefore, about 3 red AGNs are expected to be detected 
in our surveys with {\it ASCA}. 
This is consistent with our observations that 
two red AGNs with $K_S<14$ mag (or four, if errors in the 
$J-K_S$ color is taken into account) are detected at $z<0.3$; 
one is a narrow-line AGN (NO07), 
and the other is a broad-line AGN (NO10). 

\section{Summary}
We present NIR colors of hard X-ray selected 
AGNs detected with {\it ASCA} surveys (ALSS, AMSSn). 
The flux limit of the sample is 
$(1-3) \times 10^{-13}$ erg s$^{-1}$ cm$^{-2}$ 
in the 2--10 keV band in a total surveyed area of $\sim$75 deg$^2$. 
Optical follow-up identifications are successful and 
virtually all the sources were identified. 
The sample covers the AGNs at 0.1 $\lesssim z \lesssim$ 1 
with $L_{\rm 2-10keV} \sim 10^{42} - 10^{46}$ erg s$^{-1}$ 
with very high completeness. 
The NIR photometric data of 46 AGNs of the sample are 
taken from 2MASS Point Source Catalog, 
49 AGNs from the observations with KPNO 2.1m telescope, 
and 9 AGNs from the observations with UH88 inch telescope. 
The $J-K_{S}$ color distribution of the {\it ASCA} AGNs mostly 
traces that of optically-selected QSOs. 
The fraction of red ($J-K_{S}>2$ mag) AGNs in our sample 
is $\sim2\pm1\%$, 
which is almost the same as that in optically-selected QSOs. 
The number of red AGNs found in our sample is 
consistent with that estimated from 
the surface density of red AGNs found in 2MASS (Cutri et al. 2001) 
and the X-ray to $K_S$ flux ratios 
for the subsample of them (Wilkes et al. 2002). 
We also found that
all QSOs ($L_{\rm 2-10~keV} \ge 10^{44.5}$ erg s$^{-1}$) with 
a large X-ray absorption ($N_{\rm H} \ge 10^{22}$ cm$^{-2}$) 
show no or very small absorption in NIR wavelength 
compared with their X-ray absorption. 
Estimated dust-to-gas ratios ($A_V/N_{\rm H}$) of them 
are smaller than the Galactic value by about one or two orders of magnitude. 
Therefore, the anomaly of dust-to-gas ratio 
in a circumnuclear region of an AGN, which has been shown 
for Seyfert-class AGNs by Maiolino et al. (2001), 
is also seen in QSO-class AGNs. 
Since a fraction of this population in our sample is 
about 30\% among the sample with the X-ray luminosity larger than
$10^{44.5}$ erg s$^{-1}$, such fraction of optical/UV-selected
type 1 QSOs known to date may have large amount of dust-less 
circumnuclear gases. 

\acknowledgments

We would like to thank the staff members of 
KPNO and UH88 inch telescope. 
This publication makes use of data products from 
the Two Micron All Sky Survey, which is a joint project of 
the University of Massachusetts and 
the Infrared Processing and Analysis Center, 
California Institute of Technology, funded by NASA and 
the National Science Foundation.

\clearpage

\clearpage

\clearpage

\clearpage

\clearpage

\clearpage

\include{tab1}

\end{document}

%% file: tab1.tex
\oddsidemargin -1cm
\textheight 21.3cm
\textwidth 15cm
\begin{deluxetable}{p{3.6cm}p{1.2cm}p{1.2cm}p{1.4cm}p{0.8cm}p{0.8cm}p{0.8cm}p{1cm}p{0.8cm}p{0.8cm}p{0.8cm}p{1.1cm}p{1cm}}
\tabletypesize{\scriptsize}
\rotate
\tablecaption{Properties of {\it ASCA} AGNs \label{tbl-1}}
\tablewidth{0pt}
\tablehead{
\multicolumn{1}{c}{Name\tablenotemark{a}}            &
\multicolumn{1}{c}{$J$} &
\multicolumn{1}{c}{$H$} &
\multicolumn{1}{c}{$K_{S}$} &
\multicolumn{1}{c}{$J\!-\!K_{S}$} & 
\multicolumn{1}{c}{$B$\tablenotemark{b}} &
\multicolumn{1}{c}{$R$\tablenotemark{b}} &
\multicolumn{1}{c}{NIR Data\tablenotemark{c}} &
\multicolumn{1}{c}{$z$\tablenotemark{d}}   & 
\multicolumn{1}{c}{flux\tablenotemark{e}}   &
\multicolumn{1}{c}{$\log L$\tablenotemark{f}}   &
\multicolumn{1}{c}{$N_{\rm H}$\tablenotemark{g}}   &
\multicolumn{1}{c}{$\Gamma$\tablenotemark{h}}   \\
\multicolumn{1}{c}{}            &
\multicolumn{1}{c}{}   &
\multicolumn{1}{c}{}   &
\multicolumn{1}{c}{}   &
\multicolumn{1}{c}{}   &
\multicolumn{1}{c}{}   &
\multicolumn{1}{c}{}   &
\multicolumn{1}{c}{Source}   & 
\multicolumn{1}{c}{}   & 
\multicolumn{1}{c}{${\rm 2\!-\!10keV}$}   & 
\multicolumn{1}{c}{${\rm 2\!-\!10keV}$}   & 
\multicolumn{1}{c}{}   &
\multicolumn{1}{c}{}   \\
\multicolumn{1}{c}{}            &
\multicolumn{1}{c}{(mag)}   &
\multicolumn{1}{c}{(mag)}   &
\multicolumn{1}{c}{(mag)}   &
\multicolumn{1}{c}{(mag)}   &
\multicolumn{1}{c}{(mag)}   &
\multicolumn{1}{c}{(mag)}   &
\multicolumn{1}{c}{}   & 
\multicolumn{1}{c}{}   & 
\multicolumn{1}{c}{}   & 
\multicolumn{1}{c}{}   & 
\multicolumn{1}{c}{}   &
\multicolumn{1}{c}{}  
}
\startdata
1AXG~J000605$+$2031~~NE01 & 16.01$\pm$0.08 & 15.59$\pm$0.13 & 14.38$\pm$0.06 & 1.63 & 16.3 & 16.2 & 2MASS & 0.385 & 7.17 & 44.59 & 0 & 2.2$_{-0.2}^{+0.3}$ \\ 
1AXG~J000927$-$0438~~NO19 & 16.99$\pm$0.17 & 16.08$\pm$0.17 & 15.34$\pm$0.16 & 1.65 & 19.3 & 18.3 & 2MASS & 0.314 & 3.44 & 44.04 & 0 & 2.1$_{-0.2}^{+0.2}$ \\ 
1AXG~J001913$+$1556~~NE18 & 16.72$\pm$0.14 & 15.81$\pm$0.13 & 15.14$\pm$0.14 & 1.58 & 17.8 & 17.5 & 2MASS & 2.270 & 3.46 & 46.01 & 0 & 1.9$_{-0.2}^{+0.2}$ \\ 
1AXG~J002619$+$1050~~NO49 & 16.46$\pm$0.13 & 15.88$\pm$0.18 & 14.93$\pm$0.15 & 1.53 & 18.0 & 17.5 & 2MASS & 0.474 & 3.05 & 44.38 & 0.37$_{-0.35}^{+0.38}$ & 1.9 \\ 
1AXG~J002637$+$1725~~NO50 & 14.78$\pm$0.07 & 14.11$\pm$0.09 & 13.62$\pm$0.06 & 1.16 & 15.9 & 14.4 & 2MASS & 0.043 & 3.12 & 42.15 & 1.34$_{-0.36}^{+0.44}$ & 1.9 \\ 
1AXG~J015840$+$0347~~NO46 & 15.57$\pm$0.05 & 15.08$\pm$0.09 & 14.19$\pm$0.06 & 1.38 & 17.0 & 16.8 & 2MASS & 0.658 & 4.05 & 44.91 & 0 & 2.3$_{-0.2}^{+0.2}$ \\ 
1AXG~J023520$-$0347~~NO16 & 16.39$\pm$0.10 & 15.44$\pm$0.11 & 14.52$\pm$0.10 & 1.87 & 17.5 & 17.4 & 2MASS & 0.376 & 4.24 & 44.34 & 0 & 2.3$_{-0.1}^{+0.2}$ \\ 
1AXG~J033516$-$1505~~SE20 & 15.83$\pm$0.08 & 14.95$\pm$0.09 & 13.90$\pm$0.06 & 1.92 & 18.1 & 16.5 & 2MASS & 0.122 & 4.80 & 43.28 & 0.58$_{-0.18}^{+0.19}$ & 1.9 \\ 
1AXG~J035137$-$1204~~SE17 & 16.09$\pm$0.10 & 15.33$\pm$0.10 & 14.80$\pm$0.12 & 1.29 & 17.6 & 17.1 & 2MASS & 0.182 & 3.42 & 43.51 & 0.80$_{-0.35}^{+0.40}$ & 1.9 \\ 
1AXG~J041757$+$0101~~NE25 & 15.82$\pm$0.03 & 15.32$\pm$0.04 & 14.52$\pm$0.02 & 1.30 & $\cdots$ & 17.0\tablenotemark{S} & KPNO & 0.126 & 3.12 & 43.12 & 0.43$_{-0.32}^{+0.39}$ & 1.9 \\ 
1AXG~J043420$-$0822~~NE03 & 15.94$\pm$0.09 & 15.28$\pm$0.10 & 14.55$\pm$0.09 & 1.39 & 18.0 & 17.9 & 2MASS & 0.154 & 14.55 & 43.98 & 0.71$_{-0.16}^{+0.19}$ & 1.9 \\ 
1AXG~J044749$-$0629~~NO08 & 16.39$\pm$0.14 & 15.36$\pm$0.14 & 14.74$\pm$0.12 & 1.66 & 18.4 & 17.7 & 2MASS & 0.213 & 3.32 & 43.64 & 0.31$_{-0.22}^{+0.24}$ & 1.9 \\ 
1AXG~J083747$+$6513~~NO06 & 16.73$\pm$0.04 & 16.29$\pm$0.07 & 15.42$\pm$0.04 & 1.31 & 18.5 & 18.1 & KPNO & 1.105 & 5.59 & 45.52 & 0 & 2.0$_{-0.1}^{+0.1}$ \\ 
1AXG~J090053$+$3856~~NO54 & 16.72$\pm$0.15 & 16.09$\pm$0.21 & $>$15.28 & 1.44 & 20.8\tablenotemark{U} & 18.2\tablenotemark{U} & 2MASS & 0.229 & 5.42 & 44.02 & 5.23$_{-1.29}^{+1.92}$ & 1.9 \\ 
1AXG~J102337$+$1936~~NE12 & 17.27$\pm$0.10 & 16.61$\pm$0.13 & 15.44$\pm$0.08 & 1.83 & 18.8 & 18.0 & KPNO & 0.407 & 3.37 & 44.29 & 1.31$_{-0.64}^{+0.76}$ & 1.9 \\ 
1AXG~J103934$+$5330~~NO11 & 16.55$\pm$0.13 & 15.91$\pm$0.17 & 15.38$\pm$0.15 & 1.17 & 18.2 & 17.6 & 2MASS & 0.229 & 3.24 & 43.69 & 0.10$_{-0.10}^{+0.26}$ & 1.9 \\ 
1AXG~J104026$+$2046~~NO41 & 17.75$\pm$0.02 & 16.87$\pm$0.05 & 16.18$\pm$0.04 & 1.57 & 20.0\tablenotemark{U} & 19.5\tablenotemark{U} & UH & 0.467 & 10.40 & 44.90 & 0.12$_{-0.12}^{+0.15}$ & 1.9 \\ 
1AXG~J105722$-$0351~~NE14 & 17.12$\pm$0.08 & 16.58$\pm$0.14 & 15.66$\pm$0.07 & 1.46 & 17.8 & 17.8 & KPNO & 0.555 & 3.15 & 44.55 & 0.41$_{-0.41}^{+0.62}$ & 1.9 \\ 
1AXG~J111432$+$4055~~NE26 & 15.81$\pm$0.08 & 15.14$\pm$0.08 & 14.25$\pm$0.08 & 1.56 & 18.8 & 16.4 & 2MASS & 0.153 & 8.35 & 43.82 & 4.10$_{-0.98}^{+1.37}$ & 1.9 \\ 
1AXG~J111518$+$4042~~NO56 & 15.23$\pm$0.06 & 14.51$\pm$0.07 & 13.98$\pm$0.06 & 1.25 & 16.6 & 15.1 & 2MASS & 0.079 & 12.73 & 43.29 & 0 & 2.0$_{-0.0}^{+0.0}$ \\ 
1AXG~J121328$+$2938~~NO28 & 16.45$\pm$0.12 & 15.68$\pm$0.13 & 15.23$\pm$0.15 & 1.22 & $\cdots$ & 16.8\tablenotemark{S} & 2MASS & 0.143 & 4.62 & 43.40 & 0 & 1.9$_{-0.1}^{+0.1}$ \\ 
1AXG~J121359$+$1404~~NO01 & 16.46$\pm$0.11 & 15.67$\pm$0.13 & 14.91$\pm$0.12 & 1.55 & 17.6 & 17.0 & 2MASS & 0.154 & 3.21 & 43.31 & 0.06$_{-0.06}^{+0.22}$ & 1.9 \\ 
1AXG~J121427$+$2936~~NO27 & 16.09$\pm$0.03 & 15.38$\pm$0.04 & 14.46$\pm$0.02 & 1.63 & 19.1 & 17.5 & KPNO & 0.309 & 7.42 & 44.34 & 0.29$_{-0.15}^{+0.16}$ & 1.9 \\ 
1AXG~J121854$+$2957~~NO07 & 16.17$\pm$0.01 & 14.73$\pm$0.01 & 13.19$\pm$0.01 & 2.98 & 20.1\tablenotemark{U} & 18.5\tablenotemark{U} & UH & 0.178 & 8.6\tablenotemark{i} & 44.04\tablenotemark{i} & 8.2$_{-0.7}^{+1.1}\tablenotemark{i}$ & 2.0$_{-0.2}^{+0.2}$\tablenotemark{i} \\ 
1AXG~J121930$+$0643~~NO44 & 14.29$\pm$0.01 & 13.62$\pm$0.02 & 12.87$\pm$0.01 & 1.42 & 16.9 & 16.0 & KPNO & 0.081 & 6.30 & 43.01 & 0 & 2.2$_{-0.1}^{+0.1}$ \\ 
1AXG~J121930$+$7532~~NO39 & 16.90$\pm$0.07 & 16.10$\pm$0.08 & 15.36$\pm$0.04 & 1.54 & 17.4 & 17.2 & KPNO & 0.464 & 3.08 & 44.42 & 0 & 2.3$_{-0.2}^{+0.2}$ \\ 
1AXG~J122003$-$0025~~NO03 & 17.14$\pm$0.05 & 16.64$\pm$0.09 & 15.94$\pm$0.05 & 1.20 & 19.3 & 18.9 & KPNO & 0.422 & 5.08 & 44.50 & 1.12$_{-0.43}^{+0.49}$ & 1.9 \\ 
1AXG~J122017$+$0641~~NO45 & 15.82$\pm$0.04 & 15.10$\pm$0.04 & 14.16$\pm$0.02 & 1.66 & 18.6\tablenotemark{S} & 17.7\tablenotemark{S} & KPNO & 0.287 & 10.53 & 44.42 & 0 & 1.9$_{-0.1}^{+0.1}$ \\ 
1AXG~J122049$+$7505~~NO37 & 17.35$\pm$0.05 & 16.69$\pm$0.07 & 16.01$\pm$0.06 & 1.34 & 18.3 & 17.7 & KPNO & 0.650 & 4.49 & 44.88 & 0 & 2.0$_{-0.2}^{+0.2}$ \\ 
1AXG~J122135$+$7518~~NO38 & 13.60$\pm$0.03 & 12.68$\pm$0.03 & 11.88$\pm$0.03 & 1.72 & 15.0 & 13.9 & 2MASS & 0.073 & 99.70 & 44.11 & 0 & 2.0$_{-0.0}^{+0.0}$ \\ 
1AXG~J122155$+$7525~~NO40 & 16.32$\pm$0.13 & 15.42$\pm$0.12 & 14.63$\pm$0.11 & 1.69 & 19.7\tablenotemark{U} & 17.4\tablenotemark{U} & 2MASS & 0.239 & 2.95 & 43.69 & 0.26$_{-0.26}^{+0.31}$ & 1.9 \\ 
1AXG~J123605$+$2613~~NE04 & 17.80$\pm$0.03 & 17.00$\pm$0.04 & 16.36$\pm$0.03 & 1.44 & 21.7 & 19.5\tablenotemark{S} & KPNO & 0.459 & 3.15 & 44.37 & 0.78$_{-0.61}^{+0.71}$ & 1.9 \\ 
1AXG~J125732$+$3543~~NO31 & 16.51$\pm$0.02 & 15.63$\pm$0.02 & 14.80$\pm$0.01 & 1.71 & 18.3 & 18.0 & UH & 0.524 & 5.10 & 44.70 & 0.07$_{-0.07}^{+0.27}$ & 1.9 \\ 
1AXG~J125812$+$3519~~NE16 & 16.83$\pm$0.15 & 15.92$\pm$0.15 & 15.13$\pm$0.14 & 1.71 & 21.0 & 18.5 & 2MASS & 0.310 & 25.22 & 44.88 & 0.14$_{-0.14}^{+0.15}$ & 1.9 \\ 
1AXG~J125828$+$3528~~NE15 & 16.48$\pm$0.11 & 15.79$\pm$0.14 & 15.25$\pm$0.13 & 1.23 & 18.5 & 17.4 & 2MASS & 1.900 & 3.26 & 45.91 & 0 & 2.1$_{-0.2}^{+0.2}$ \\ 
1AXG~J130407$+$3533~~NO30 & 16.11$\pm$0.09 & 15.37$\pm$0.10 & 14.37$\pm$0.07 & 1.74 & 19.2 & 17.6 & 2MASS & 0.329 & 10.29 & 44.54 & 0.15$_{-0.11}^{+0.11}$ & 1.9 \\ 
1AXG~J130453$+$3548~~NO29 & 17.19$\pm$0.07 & 16.93$\pm$0.11 & 15.87$\pm$0.06 & 1.32 & 20.8 & 18.9 & KPNO & 0.316 & 3.13 & 44.01 & 1.28$_{-0.62}^{+0.74}$ & 1.9 \\ 
1AXG~J132310$-$1656~~SE34 & 13.14$\pm$0.08 & 12.29$\pm$0.09 & 11.94$\pm$0.07 & 1.20 & $\cdots$ & 15.0\tablenotemark{S} & 2MASS & 0.022 & 4.44 & 41.69 & 0.19$_{-0.19}^{+0.25}$ & 1.9 \\ 
1AXG~J134412$+$0016~~NE20 & 16.82$\pm$0.08 & 15.95$\pm$0.09 & 15.17$\pm$0.06 & 1.65 & $\cdots$ & 19.0 & KPNO & 0.452 & 3.69 & 44.42 & 0.56$_{-0.49}^{+0.56}$ & 1.9 \\ 
1AXG~J134450$+$0005~~NE19 & 15.42$\pm$0.08 & 14.57$\pm$0.09 & 13.90$\pm$0.09 & 1.53 & 16.9 & 15.6 & 2MASS & 0.087 & 4.85 & 42.96 & 0 & 2.2$_{-0.1}^{+0.1}$ \\ 
1AXG~J134741$-$1122~~SE30 & 15.48$\pm$0.07 & 14.72$\pm$0.04 & 14.27$\pm$0.06 & 1.21 & 16.5 & 17.2 & KPNO & 0.100 & 6.38 & 43.22 & 0.47$_{-0.32}^{+0.37}$ & 1.9 \\ 
1AXG~J140528$+$2224~~NO15 & 16.00$\pm$0.07 & 15.28$\pm$0.09 & 14.48$\pm$0.07 & 1.43 & 17.8 & 16.5 & 2MASS & 0.156 & 4.59 & 43.49 & 0 & 2.1$_{-0.1}^{+0.1}$ \\ 
1AXG~J140532$+$5055~~NO13 & 15.28$\pm$0.01 & 14.48$\pm$0.02 & 13.87$\pm$0.01 & 1.41 & $\cdots$ & 17.8\tablenotemark{S} & KPNO & 0.106 & 3.00 & 42.93 & 0 & 2.0$_{-0.2}^{+0.2}$ \\ 
1AXG~J141240$-$1209~~SE03 & 15.78$\pm$0.07 & 14.90$\pm$0.07 & 13.89$\pm$0.02 & 1.89 & 17.0 & 16.7 & KPNO & 0.247 & 23.26 & 44.62 & 0.13$_{-0.06}^{+0.07}$ & 1.9 \\ 
1AXG~J141426$-$1209~~SE28 & 18.24$\pm$0.04 & 17.58$\pm$0.04 & 17.11$\pm$0.05 & 1.13 & 18.8 & 18.4 & UH & 1.156 & 4.80 & 45.47 & 0 & 1.9$_{-0.2}^{+0.2}$ \\ 
1AXG~J142353$+$2247~~NE09 & 16.06$\pm$0.04 & 15.28$\pm$0.05 & 14.55$\pm$0.05 & 1.51 & 18.5 & 16.8 & KPNO & 0.282 & 9.08 & 44.34 & 0.10$_{-0.10}^{+0.18}$ & 1.9 \\ 
1AXG~J142651$+$2619~~NO51 & 16.40$\pm$0.13 & 15.54$\pm$0.12 & 14.94$\pm$0.11 & 1.46 & 19.6 & 17.6 & 2MASS & 0.258 & 7.33 & 44.16 & 0 & 1.9$_{-0.1}^{+0.1}$ \\ 
1AXG~J144055$+$5204~~NE13 & 16.26$\pm$0.12 & 15.77$\pm$0.16 & 14.47$\pm$0.10 & 1.79 & 17.3 & 16.7 & 2MASS & 0.320 & 5.43 & 44.28 & 0 & 2.3$_{-0.2}^{+0.2}$ \\ 
1AXG~J144109$+$3520~~NO32 & 15.56$\pm$0.07 & 14.90$\pm$0.10 & 14.54$\pm$0.08 & 1.02 & 16.7 & 15.8 & 2MASS & 0.077 & 3.11 & 42.65 & 0 & 2.0$_{-0.2}^{+0.2}$ \\ 
1AXG~J144301$+$5208~~NO26 & 16.22$\pm$0.11 & 15.61$\pm$0.16 & 14.61$\pm$0.09 & 1.61 & 19.4 & 17.2 & 2MASS & 0.212 & 3.58 & 43.69 & 1.32$_{-0.34}^{+0.41}$ & 1.9 \\ 
1AXG~J150339$+$1016~~NO05 & 15.12$\pm$0.03 & 14.45$\pm$0.04 & 14.01$\pm$0.04 & 1.11 & 17.6 & 16.1 & KPNO & 0.095 & 7.51 & 43.29 & 1.95$_{-0.26}^{+0.27}$ & 1.9 \\ 
1AXG~J150423$+$1029~~NO04 & 15.42$\pm$0.01 & 14.44$\pm$0.01 & 13.47$\pm$0.01 & 1.95 & 19.6 & 19.5 & KPNO & 1.839 & 6.54 & 46.08 & 5.41$_{-1.65}^{+1.82}$ & 1.9 \\ 
1AXG~J150430$+$4741~~NO12 & 17.43$\pm$0.06 & 17.20$\pm$0.10 & 16.55$\pm$0.09 & 0.88 & 19.3 & 18.5 & KPNO & 0.822 & 2.98 & 44.95 & 0 & 2.0$_{-0.1}^{+0.1}$ \\ 
1AXG~J151441$+$3650~~NE10 & 15.34$\pm$0.02 & 14.63$\pm$0.04 & 13.88$\pm$0.02 & 1.46 & 15.9 & 15.8 & KPNO & 0.371 & 19.61 & 44.94 & 0.11$_{-0.11}^{+0.16}$ & 1.9 \\ 
1AXG~J151524$+$3639~~NO21 & 16.68$\pm$0.03 & 15.83$\pm$0.04 & 15.07$\pm$0.03 & 1.61 & $\cdots$ & 18.5 & KPNO & 0.324 & 3.14 & 44.03 & 0.90$_{-0.49}^{+0.55}$ & 1.9 \\ 
1AXG~J155810$+$6401~~NO22 & 17.10$\pm$0.05 & 16.38$\pm$0.06 & 15.60$\pm$0.05 & 1.50 & $\cdots$ & 19.1 & KPNO & 0.352 & 5.78 & 44.36 & 0.15$_{-0.15}^{+0.21}$ & 1.9 \\ 
1AXG~J160118$+$0844~~NO53 & 17.96$\pm$0.11 & 16.83$\pm$0.10 & 16.03$\pm$0.06 & 1.93 & $\cdots$ & 20.3\tablenotemark{S} & KPNO & 0.606 & 4.00 & 44.77 & 2.95$_{-1.23}^{+1.58}$ & 1.9 \\ 
1AXG~J163538$+$3809~~NO47 & 16.25$\pm$0.03 & 15.54$\pm$0.03 & 15.19$\pm$0.04 & 1.06 & 19.3 & 17.5 & KPNO & 0.099 & 4.27 & 43.03 & 0.38$_{-0.15}^{+0.17}$ & 1.9 \\ 
1AXG~J163720$+$8207~~NE08 & 13.52$\pm$0.01 & 12.89$\pm$0.02 & 12.56$\pm$0.03 & 0.96 & $\cdots$ & 16.0\tablenotemark{S} & KPNO & 0.041 & 22.62 & 42.93 & 0.20$_{-0.11}^{+0.11}$ & 1.9 \\ 
1AXG~J170305$+$4526~~NO35 & 15.54$\pm$0.06 & 14.69$\pm$0.07 & 13.67$\pm$0.05 & 1.88 & 17.3 & 16.8 & 2MASS & 0.171 & 6.66 & 43.73 & 0.26$_{-0.14}^{+0.15}$ & 1.9 \\ 
1AXG~J170548$+$2412~~NE11 & 15.23$\pm$0.01 & 14.41$\pm$0.01 & 13.75$\pm$0.01 & 1.48 & 16.5\tablenotemark{U} & 14.5\tablenotemark{U} & KPNO & 0.114 & 5.93 & 43.29 & 0.00$_{-0.00}^{+0.26}$ & 1.9 \\ 
1AXG~J170730$+$2353~~NO24 & 17.48$\pm$0.09 & 17.05$\pm$0.11 & 16.20$\pm$0.10 & 1.28 & $\cdots$ & 20.0 & KPNO & 0.245 & 4.20 & 44.00 & 7.07$_{-2.90}^{+41.91}$ & 1.9 \\ 
1AXG~J171125$+$7111~~NO02 & 17.97$\pm$0.08 & 17.42$\pm$0.09 & 16.63$\pm$0.08 & 1.34 & 21.2 & 18.5\tablenotemark{S} & KPNO & 1.011 & 3.52 & 45.31 & 0 & 2.2$_{-0.1}^{+0.1}$ \\ 
1AXG~J171811$+$6727~~NE21 & 17.09$\pm$0.06 & 16.40$\pm$0.05 & 15.63$\pm$0.06 & 1.46 & 17.8 & 17.7 & KPNO & 0.549 & 4.04 & 44.66 & 1.51$_{-0.55}^{+0.60}$ & 1.9 \\ 
1AXG~J172938$+$5230~~NO10 & 15.72$\pm$0.07 & 14.86$\pm$0.06 & 13.69$\pm$0.05 & 2.03 & 16.4 & 16.2 & 2MASS & 0.278 & 6.49 & 44.22 & 0 & 2.3$_{-0.1}^{+0.1}$ \\ 
1AXG~J174652$+$6836~~NO42 & 14.69$\pm$0.07 & 13.84$\pm$0.08 & 12.83$\pm$0.05 & 1.86 & $\cdots$ & 16.0\tablenotemark{S} & 2MASS & 0.063 & 21.46 & 43.31 & 0 & 2.0$_{-0.0}^{+0.0}$ \\ 
1AXG~J174943$+$6823~~NO43 & 14.42$\pm$0.09 & 13.53$\pm$0.08 & 12.99$\pm$0.05 & 1.43 & $\cdots$ & 16.0\tablenotemark{S} & 2MASS & 0.051 & 9.86 & 42.81 & 0.75$_{-0.18}^{+0.18}$ & 1.9 \\ 
1AXG~J210738$-$0512~~NO17 & 19.42$\pm$0.10 & 18.52$\pm$0.08 & 17.43$\pm$0.15 & 1.99 & $\cdots$ & 20.6\tablenotemark{S} & UH & 0.841 & 3.90 & 45.09 & 4.27$_{-1.67}^{+2.10}$ & 1.9 \\ 
1AXG~J230738$-$1526~~SE14 & 17.18$\pm$0.21 & 16.14$\pm$0.21 & 15.25$\pm$0.15 & 1.93 & 19.4 & 18.3 & 2MASS & 0.199 & 3.61 & 43.60 & 0.03$_{-0.03}^{+0.24}$ & 1.9 \\ 
1AXG~J232639$+$2205~~NO48 & 15.98$\pm$0.14 & 15.32$\pm$0.11 & 14.28$\pm$0.09 & 1.70 & 18.6\tablenotemark{U} & 17.2\tablenotemark{U} & 2MASS & 0.151 & 3.77 & 43.37 & 0.31$_{-0.31}^{+0.34}$ & 1.9 \\ 
1AXG~J233200$+$1945~~NO18 & 19.44$\pm$0.13 & 18.17$\pm$0.12 & 17.70$\pm$0.10 & 1.74 & $\cdots$ & 21.8\tablenotemark{U} & UH & 1.416 & 3.31 & 45.55 & 10.86$_{-4.52}^{+6.25}$ & 1.9 \\ 
1AXG~J233253$+$1513~~NO14 & 15.91$\pm$0.09 & 15.14$\pm$0.09 & 14.21$\pm$0.09 & 1.70 & 18.0 & 16.6 & 2MASS & 0.215 & 11.84 & 44.19 & 0 & 1.9$_{-0.1}^{+0.1}$ \\ 
1AXG~J234725$+$0053~~NE23 & 15.97$\pm$0.01 & 15.20$\pm$0.01 & 14.17$\pm$0.01 & 1.80 & 19.5 & 17.5 & UH & 0.213 & 7.20 & 44.08 & 1.42$_{-0.35}^{+0.43}$ & 1.9 \\ 
1AXG~J235554$+$2836~~NO20 & 16.14$\pm$0.09 & 15.73$\pm$0.14 & 14.61$\pm$0.07 & 1.53 & 18.1 & 17.7 & 2MASS & 0.729 & 5.39 & 45.06 & 0.73$_{-0.53}^{+0.57}$ & 1.9 \\ 
AX~J130826$+$3005~~LSS023 & 15.28$\pm$0.05 & 15.17$\pm$0.08 & 14.28$\pm$0.07 & 1.00 & 17.1 & 16.3 & 2MASS & 0.803 & 3.48 & 44.96 & 0 & 1.9$_{-0.2}^{+0.2}$ \\ 
AX~J130840$+$2955~~LSS014 & 16.59$\pm$0.14 & 15.93$\pm$0.16 & 15.61$\pm$0.21 & 0.98 & 19.2 & 17.3 & 2MASS & 0.164 & 1.93 & 43.21 & 2.64$_{-1.27}^{+1.27}$ & 1.9 \\ 
AX~J130926$+$2952~~LSS016 & 18.13$\pm$0.10 & 17.13$\pm$0.06 & 16.56$\pm$0.06 & 1.57 & 21.9\tablenotemark{U} & 19.9\tablenotemark{U} & KPNO & 0.375 & 1.81 & 43.96 & 2.82$_{-1.25}^{+1.90}$ & 1.9 \\ 
AX~J131015$+$3004~~LSS033 & 17.15$\pm$0.07 & 16.49$\pm$0.10 & 16.26$\pm$0.08 & 0.89 & 18.9 & 17.9 & KPNO & 1.165 & 1.92 & 45.08 & 0.00$_{-0.00}^{+1.92}$ & 1.9 \\ 
AX~J131021$+$3019~~LSS039 & 18.91$\pm$0.11 & 17.84$\pm$0.09 & 16.91$\pm$0.08 & 2.00 & 21.6\tablenotemark{U} & 19.4\tablenotemark{K} & KPNO & 1.152\tablenotemark{j} & 2.62\tablenotemark{k} & 45.25\tablenotemark{k} & 0.00$_{-0.00}^{+0.02}$\tablenotemark{k} & 2.0$_{-0.1}^{+0.1}$\tablenotemark{k} \\ 
AX~J131054$+$3004~~LSS037 & 18.73$\pm$0.06 & $\cdots$ & 17.95$\pm$0.09 & 0.78 & 19.9 & 19.2 & UH & 1.577 & 3.05 & 45.59 & 2.48$_{-1.71}^{+1.71}$ & 1.9 \\ 
AX~J131128$+$3105~~LSS080 & 16.44$\pm$0.13 & 16.24$\pm$0.23 & 15.39$\pm$0.18 & 1.04 & 17.7 & 17.1 & 2MASS & 1.331 & 1.42 & 45.08 & 0.00$_{-0.00}^{+2.88}$ & 1.9 \\ 
AX~J131156$+$3054~~LSS075 & 17.72$\pm$0.10 & 16.97$\pm$0.11 & 15.86$\pm$0.07 & 1.86 & 20.4 & 19.2 & KPNO & 0.366 & 1.76 & 43.88 & 0.00$_{-0.00}^{+0.83}$ & 1.9 \\ 
AX~J131210$+$3048~~LSS072 & 16.37$\pm$0.13 & $>$15.36 & 14.90$\pm$0.12 & 1.47 & 20.5\tablenotemark{U} & 18.4\tablenotemark{U} & 2MASS & 0.189 & 1.79 & 43.28 & 1.42$_{-0.70}^{+0.99}$ & 1.9 \\ 
AX~J131249$+$3112~~LSS096 & 16.97$\pm$0.09 & 17.01$\pm$0.21 & 16.13$\pm$0.09 & 0.83 & 18.6 & 17.9 & KPNO & 1.055 & 3.43 & 45.24 & 1.00$_{-0.60}^{+0.60}$ & 1.9 \\ 
AX~J131321$+$3119~~LSS103 & 18.43$\pm$0.06 & 17.86$\pm$0.08 & 16.96$\pm$0.06 & 1.47 & 19.6 & 19.8 & KPNO & 0.834 & 1.16 & 44.53 & 0.92$_{-0.65}^{+0.65}$ & 1.9 \\ 
AX~J131327$+$3155~~LSS121 & 16.71$\pm$0.04 & 15.83$\pm$0.07 & 15.19$\pm$0.04 & 1.52 & $\cdots$ & 17.6\tablenotemark{K} & KPNO & 0.303 & 2.07 & 43.77 & 0.16$_{-0.16}^{+0.24}$ & 1.9 \\ 
AX~J131345$+$3118~~LSS104 & 17.80$\pm$0.04 & 17.35$\pm$0.10 & 16.55$\pm$0.06 & 1.25 & $\cdots$ & 19.3\tablenotemark{K} & KPNO & 0.789 & 3.27 & 44.92 & 0.16$_{-0.16}^{+0.39}$ & 1.9 \\ 
AX~J131407$+$3158~~LSS127 & 15.38$\pm$0.02 & 14.81$\pm$0.03 & 14.14$\pm$0.02 & 1.24 & 18.1 & 16.7 & KPNO & 0.177 & 1.91 & 43.21 & 0.07$_{-0.07}^{+0.27}$ & 1.9 \\ 
AX~J131501$+$3141~~LSS119 & 13.79$\pm$0.01 & 12.95$\pm$0.01 & 12.46$\pm$0.01 & 1.33 & 17.3\tablenotemark{U} & 15.6\tablenotemark{U} & KPNO & 0.072 & 2.48\tablenotemark{l} & 42.62\tablenotemark{l} & 6.40$_{-1.40}^{+1.90}$\tablenotemark{l} & 1.5$_{-0.4}^{+0.4}$\tablenotemark{l} \\ 
AX~J131521$+$3159~~LSS136 & 16.40$\pm$0.03 & 16.00$\pm$0.03 & 15.30$\pm$0.03 & 1.10 & $\cdots$ & 17.7\tablenotemark{B} & KPNO & 0.838 & 2.72 & 44.91 & 0 & 1.9$_{-0.2}^{+0.2}$ \\ 
AX~J131529$+$3117~~LSS110 & 17.16$\pm$0.07 & 16.43$\pm$0.12 & 15.64$\pm$0.07 & 1.52 & 18.9 & 18.8 & KPNO & 0.449 & 1.95 & 44.15 & 0 & 2.0$_{-0.2}^{+0.2}$ \\ 
AX~J131551$+$3237~~LSS171 & 16.36$\pm$0.03 & 15.67$\pm$0.04 & 14.93$\pm$0.03 & 1.43 & 19.7\tablenotemark{U} & 18.1\tablenotemark{U} & KPNO & 0.128 & 2.26\tablenotemark{l} & 43.03\tablenotemark{l} & 1.90$_{-0.60}^{+0.80}$\tablenotemark{l} & 1.9$_{-0.4}^{+0.4}$\tablenotemark{l} \\ 
AX~J131639$+$3149~~LSS137 & 17.90$\pm$0.08 & 17.22$\pm$0.11 & 16.43$\pm$0.07 & 1.47 & 20.0 & 20.0 & KPNO & 0.622 & 1.50 & 44.35 & 0.35$_{-0.35}^{+0.49}$ & 1.9 \\ 
AX~J131707$+$3237~~LSS175 & 18.38$\pm$0.12 & 18.21$\pm$0.22 & 18.08$\pm$0.94 & 0.30 & 19.5 & 18.6 & KPNO & 0.745 & 1.66 & 44.59 & 0 & 2.0$_{-0.2}^{+0.2}$ \\ 
AX~J131724$+$3203~~LSS152 & 17.88$\pm$0.05 & 17.61$\pm$0.07 & 16.99$\pm$0.10 & 0.89 & 19.7 & 19.1 & KPNO & 0.849 & 1.81 & 44.75 & 1.90$_{-0.99}^{+0.99}$ & 1.9 \\ 
AX~J131725$+$3300~~LSS192 & 15.83$\pm$0.08 & 14.98$\pm$0.09 & 14.34$\pm$0.07 & 1.49 & 18.9 & 16.6 & 2MASS & 0.126 & 1.48 & 42.80 & 0.42$_{-0.24}^{+0.24}$ & 1.9 \\ 
AX~J131758$+$3257~~LSS195 & 16.35$\pm$0.11 & 15.49$\pm$0.09 & 14.68$\pm$0.09 & 1.67 & 19.8\tablenotemark{U} & 18.0\tablenotemark{U} & 2MASS & 0.214 & 1.90 & 43.40 & 0.28$_{-0.19}^{+0.19}$ & 1.9 \\ 
AX~J131805$+$3349~~LSS233 & 14.92$\pm$0.06 & 14.24$\pm$0.07 & 13.76$\pm$0.05 & 1.16 & 13.8 & 13.4 & 2MASS & 0.035 & 4.83 & 42.13 & 0.28$_{-0.17}^{+0.17}$ & 1.9 \\ 
AX~J131816$+$3240~~LSS183 & 16.67$\pm$0.13 & 16.00$\pm$0.14 & 15.64$\pm$0.20 & 1.03 & 18.0\tablenotemark{U} & 17.5\tablenotemark{U} & 2MASS & 1.649 & 2.04\tablenotemark{k} & 45.56\tablenotemark{k} & 0.00$_{-0.00}^{+0.08}$\tablenotemark{k} & 2.1$_{-0.1}^{+0.1}$\tablenotemark{k} \\ 
AX~J131822$+$3347~~LSS235 & 15.48$\pm$0.06 & 14.73$\pm$0.07 & 13.87$\pm$0.05 & 1.61 & 17.3 & 16.4 & 2MASS & 0.119 & 10.14 & 43.57 & 0 & 2.0$_{-0.1}^{+0.1}$ \\ 
AX~J131831$+$3320~~LSS212 & 17.45$\pm$0.04 & 16.66$\pm$0.05 & 16.07$\pm$0.05 & 1.38 & 20.4 & 19.7 & KPNO & 0.535 & 1.90 & 44.30 & 0.36$_{-0.24}^{+0.24}$ & 1.9 \\ 
AX~J131831$+$3341~~LSS228 & 18.24$\pm$0.03 & 17.32$\pm$0.03 & 16.39$\pm$0.14\tablenotemark{m} & 1.86 & 21.7\tablenotemark{U} & 20.6\tablenotemark{U} & UH & 0.653 & 4.61 & 44.89 & 1.14$_{-0.43}^{+0.43}$ & 1.9 \\ 
AX~J131927$+$3343~~LSS238 & 17.80$\pm$0.08 & 16.78$\pm$0.08 & 16.00$\pm$0.06 & 1.80 & 19.5\tablenotemark{U} & 18.9\tablenotemark{U} & KPNO & 0.416 & 1.71 & 44.00 & 0.64$_{-0.43}^{+0.43}$ & 1.9 \\ 
AX~J131928$+$3251~~LSS197 & 17.34$\pm$0.09 & 16.67$\pm$0.10 & 15.81$\pm$0.05 & 1.53 & 19.4 & 18.4 & KPNO & 0.311 & 1.19 & 43.55 & 0.28$_{-0.28}^{+0.37}$ & 1.9 \\ 
\enddata
\tablenotetext{a}{X-ray source name and identification number 
(Ueda et al. 1999, 2001, Akiyama et al. 2000a, 2003).}
\tablenotetext{b}{B and R magnitudes from APM catalog, 
unless otherwise noted. Magnitudes are converted from 
O and E magnitudes of APM catalogue by using 
the equations by Evans (1989). 
U : Determined by photometric observations with 
a CCD camera on the UH88 inch telescope (Akiyama et al. 2003). 
S : Determined from the spectrophotometric data 
(Akiyama et al. 2003). 
K : Determined by relative photometry 
with mosaic CCD image taken at KISO observatory (Akiyama et al. 2000a). 
B : APM magnitude data is blended. Deblending was done 
using CCD image taken during the spectroscopic observations 
(Akiyama et al. 2000a). }
\tablenotetext{c}{KPNO, 2MASS, and UH shows 
Kitt Peak National Observatory (2.1m telescope), 
2MASS All-Sky Point Source Catalog, 
and University of Hawaii (88 inch telescope), respectively.}
\tablenotetext{d}{Redshift determined by the spectroscopic 
observations in optical (Akiyama et al. 2000a, 2003).}
\tablenotetext{e}{2--10keV flux in units of 10$^{-13}$ erg s$^{-1}$ cm$^{-2}$. 
Conversions from countrates are done based on the best-fit 
absorbed power-law spectra. 
(More details for the conversions are described by 
Ueda et al. 2003).}
\tablenotetext{f}{Absorption-corrected luminosity 
in the rest-frame 2--10keV band in units of erg s$^{-1}$. }
\tablenotetext{g}{Best fit rest-frame hydrogen 
column density for the absorbed power-law model 
in units of 10$^{22}$ cm$^{-2}$. 
The value of 0 without error indicates that 
$N_{\rm H}=0$ is assumed if the best-fit photon index is larger than 1.9. 
See Section 2 and Ueda et al. (2003) for the details.}
\tablenotetext{h}{Intrinsic photon index. 
The value of 1.9 without error indicates that 
the intrinsic photon index of 1.9 is assumed 
if the best-fit photon index is smaller than 1.9.
See Section 2 and Ueda et al. (2003) for the details.} 
\tablenotetext{i}{X-ray data from Loaring et al.(2003).}
\tablenotetext{j}{The identification of this X-ray source was 
not correct in Akiyama et al. (2000a). See the footnote in section 4.3.2.}
\tablenotetext{k}{X-ray data from 
the follow-up observation with {\it XMM-Newton}. See Ueda et al. (2003).}
\tablenotetext{l}{X-ray data from 
the follow-up observation with {\it ASCA}.}
\tablenotetext{m}{$K_S$ magnitude converted from K magnitude 
by Akiyama and Ohta (2001).}
\end{deluxetable}